\documentclass[]{aa}
\usepackage{graphicx}
\usepackage{txfonts}
\newcommand{\ltsima} {$\; \buildrel < \over \sim \;$}
\newcommand{\gtsima} {$\; \buildrel > \over \sim \;$}
\newcommand{\lta} {\lower.5ex\hbox{\ltsima}}
\newcommand{\gta} {\lower.5ex\hbox{\gtsima}}

\begin{document}
\title{In search of dying radio sources in the local universe}
\author{P. Parma\inst{1} \and M. Murgia\inst{1,3} \and H.R. de Ruiter\inst{2}
\and R. Fanti\inst{1} \and K.-H. Mack \inst{1}
\and F. Govoni\inst{3}}

\institute {INAF - Istituto di Radioastronomia, Via Gobetti 101, I-40129 Bologna, Italy
\and INAF - Osservatorio Astronomico di Bologna, Via Ranzani 1, I-40127 Bologna, Italy
\and INAF - Osservatorio Astronomico di Cagliari, Loc. Poggio dei Pini, Strada 54,
I-09012 Capoterra, Cagliari, Italy 
}
\offprints{P. \, Parma}
\date{Received; accepted}
\titlerunning{Dying sources}
\authorrunning{Parma et al.}
\abstract{} {Up till now very few dying sources were known, presumably because the
dying phase is short at centimeter wavelengths. We therefore have tried to improve the statistics on sources that have ceased to be active,
or are intermittently active. The latter sources would partly consist of a fossil radio plasma left over from an earlier phase of
activity, plus a recently restarted core and radio jets. Improving the statistics of dying sources will give us a better handle on the
evolution of radio sources, in particular the frequency and time scales of radio activity. 
}
{We have used the WENSS and NVSS surveys, in order to find sources with steep spectral indices, associated with nearby 
elliptical galaxies. In the cross correlation we presently used only unresolved sources, with flux densities at 1.4 GHz larger than 10 mJy.
The eleven candidates thus obtained were observed with the VLA in various configurations, in order to confirm the steepness
of the spectra, and to check whether active structures like flat-spectrum cores and jets are present, perhaps at low levels.
We estimated the duration of the active and relic phases by modelling the integrated radio spectra using the standard models of spectral
evolution. }
{We have found six dying sources and three restarted sources, while the remaining two candidates remain unresolved also with the new
VLA data and may be Compact Steep Spectrum sources, with an unusually steep spectrum.
The typical age of the active phase, as derived by spectral fits, is in the range $10^7 - 10^8$ years. For our sample of dying sources, the age
of the relic phase is on average shorter by an order of magnitude than the active phase. 
}
{}

\maketitle

\section{Introduction}
\label{sec:intro}

Radio sources associated with elliptical galaxies are supplied with energy from 
active galactic nuclei  via plasma jets or beams (Scheuer \cite{Scheuer74}; Begelman et al. 
\cite{Beg84}; Bridle \& Perley  \cite{Bridle84}).
In the active stage the total spectra of the radio sources are usually well
approximated by a power law over a wide range of frequencies; spectral 
breaks at high frequencies, with moderate steepening of the 
spectrum, are also often observed. The break frequency is related
to the age of the source.

After this phase, which may last several  $10^{7}$ years, the 
activity in the nuclei stops or falls to such a low level that  the plasma 
outflow can no longer  be sustained.
The radio core, well-defined jets and compact hot-spots will
disappear, because these structures are produced by continuous activity. 
However, the radio lobes may still remain detectable  for a long time if 
they are subject only to radiative losses. 

The switch-off of the injection of energetic electrons leads 
to a second high frequency break with an exponential steepening of the 
total spectrum of the radio source (Komissarov \& Gubanov \cite{Kom94}). 
This second high frequency break is related to the time passed since the switch-off
of the central engine. Therefore  ``dying'' radio galaxies should have steeper high frequency 
spectra than normal (active) radio galaxies. 
Current models of radio source physics, based on the assumption that radiative losses are dominant, 
predict the existence of a large population of radio 
sources that are bright at 100 MHz, but have a steep spectrum around 1 GHz.
However, such sources are not commonly found.

According to Giovannini et al. (\cite{Giov88}) only few percent  of objects in a sample of B2
and 3C radio galaxies have the characteristics of a dying radio galaxy. 
The first unambiguous example of such a source was given by Cordey (\cite{Cordey87}); see also
more recent data on this source in Jamrozy et al. (\cite{Jamrozy04}).
Up till now only four other sources have been discovered (e.g. Venturi et al. \cite{Venturi98}; 
Murgia et al. \cite{Murgia05}). 

In the literature some possible dying sources, located in rich clusters, are mentioned, but these
can be interpreted either as remnants of dying radio galaxies at the point of being disrupted by 
a particularly turbulent environment (Slee et al. \cite{Slee01}), or as old radio lobes that are 
compressed by shock waves (En\ss lin \& Gopal-Krishna \cite{Ens01}).
However the identification of former host galaxies is problematic and thus their association with
dying radio sources is uncertain.

It is also possible that radio galaxies may be active intermittently; in 
that case one may find fossil radio plasma left over from an 
earlier phase of activity, while newly restarted core and radio jets are visible as well. 
Examples of restarting activity in the presence of fossil radio lobes are 3C\,338 
and 3C\,317 (Roettiger et al. \cite{Roet93}; Venturi et al. \cite{Venturi04}); intermittent
jet activity in 4C29.30 is discussed by Jamrozy et al. (\cite{Jamrozy07}). Restarting activity has
now also been found in Compact Steep Spectrum (CSS) sources (Marecki et al. \cite{Marecki06}).

The "double-double radio 
sources" form an interesting class of restarting activity, in which new radio lobes can be
seen close to the nucleus, while an older pair of lobes is still present farther out (Schoenmaker et al. 
\cite{Schoenmaker00}). At present about a dozen of such sources are known (see Jamrozy et al. \cite{Jamrozy07}
and references therein).

Dying radio sources and intermittent radio galaxies are more easily 
detectable at low frequencies:  a search for this type of object 
should be performed using surveys at frequencies below 1 GHz.

In this paper we describe a search  for dying sources  and  discuss the properties
of the sources we have  found so far. 

In Sect. \ref{sec:sample} we briefly describe the selection of the sample and the radio and optical
observations.
In Sect. \ref{sec:results} we show radio contour maps and integrated spectra of the sources in the sample.
In Sect. \ref{sec:comments} we give some comments on individual sources and in Sect. \ref{sec:discussion}
we discuss the properties of the dying sources. Finally in Sect. \ref{sec:summary} we summarize the results
obtained in this paper.

All intrinsic parameters (radio power, absolute magnitudes, sizes) were calculated with
$H_0 = 71 $ km s$^{-1}$ Mpc$^{-1}$, $\Omega_{m} =0.3$, $\Omega_{\Lambda}=0.7$.
Spectral indices $\alpha$ are defined according to $ S \propto \nu ^{-\alpha}$. 

\section{Selection of the sample and observations}
\label{sec:sample}

\subsection{The sample}
\label{subs:sample}
Since dying radio galaxies and intermittent radio sources are more easily  detected at low frequencies, the Westerbork Northern
Sky Survey (WENSS; Rengelink et al. \cite{Renge97}) at 325 MHz is particularly well-suited for this purpose.
As a starting point we have used the work done by De Breuck et al. (\cite{debreuck00}); while their aim was to find  high-redshift 
radio galaxies, we are instead interested in steep spectrum radio sources identified with bright galaxies:
these are our prime dying source candidates. 
De Breuck et al. (\cite{debreuck00}) correlated the WENSS and NVSS 
(Condon et al. \cite{Condon98}) catalogues using a search radius of 10 arcsec centred on the WENSS positions. They  obtained 
spectral indices for 143000 sources with $S_{325MHz} \ge 30$ mJy. Only unresolved sources (angular cutoff $\sim 1$ arcmin) 
were considered. A further condition was $S_{1400} > 10 $ mJy. We note that especially this last criterion may lead to the loss of
many steep spectrum sources, because De Breuck et al. (\cite{debreuck00}) effectively used only that part of WENSS with flux densities 
above $\sim 100$ mJy. Thus they obtained a sample of 343 sources with spectral indices $ \alpha_{0.33}^{1.4} > 1.3$. We plan to
extend in the near future the search for dying radio sources also to 1400 MHz flux densities below 10 mJy. 
Optical identifications were done using the digitized POSS-I and were listed in their table A.4.
We checked all the identifications listed and estimated galaxy magnitudes following the procedure described in de Ruiter et al. 
(\cite {deruiter98}). Selecting from list A.4 only sources associated with an elliptical galaxy with $m_{r} < 17$ we obtained 11 
candidate dying sources.

\begin{figure*}
  \centering
  \includegraphics[width=18cm]{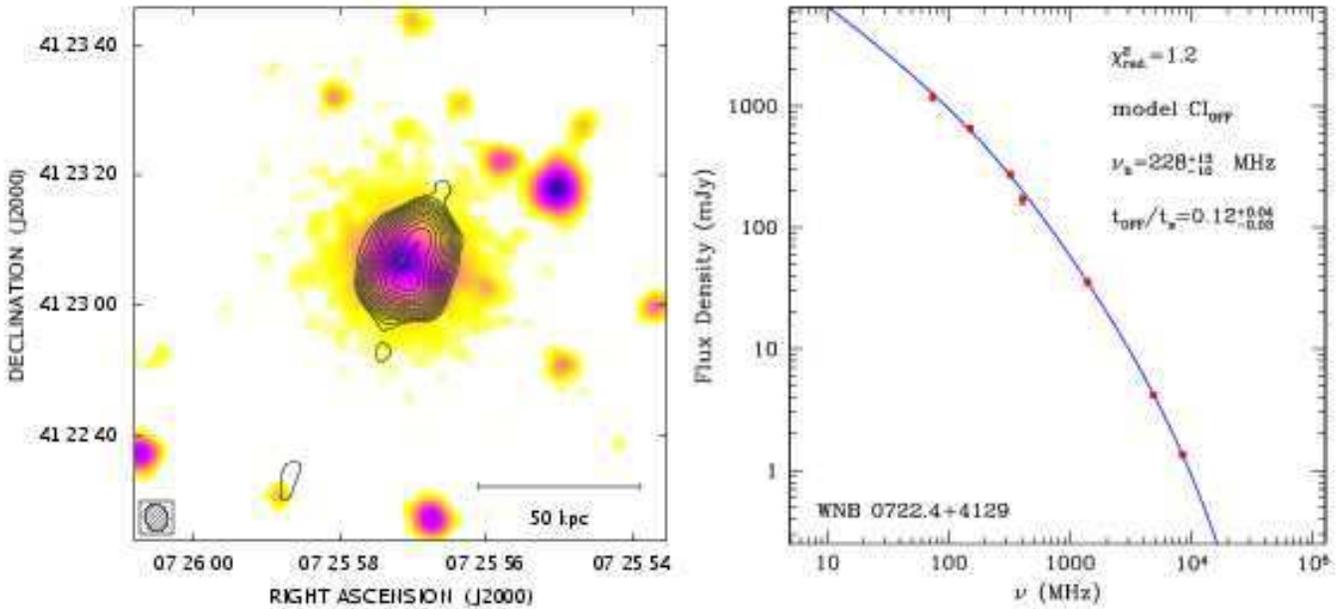}
  \caption{WNB0722.4+4129. Left: radio contours at 1.4 GHz overlayed on a DSS2 optical image. Contour levels start at 
 126 $\mu$Jy/beam (3$\sigma$) and scale by $\sqrt{2}$. The beam is $4\farcs 3\times 3\farcs 7$. 
Right: integrated radio spectrum. The solid line represents the best fit of the synchrotron model described in the text.}
  \label{fig:fig1}
\end{figure*}

\begin{figure*}
  \centering
  \includegraphics[width=18cm]{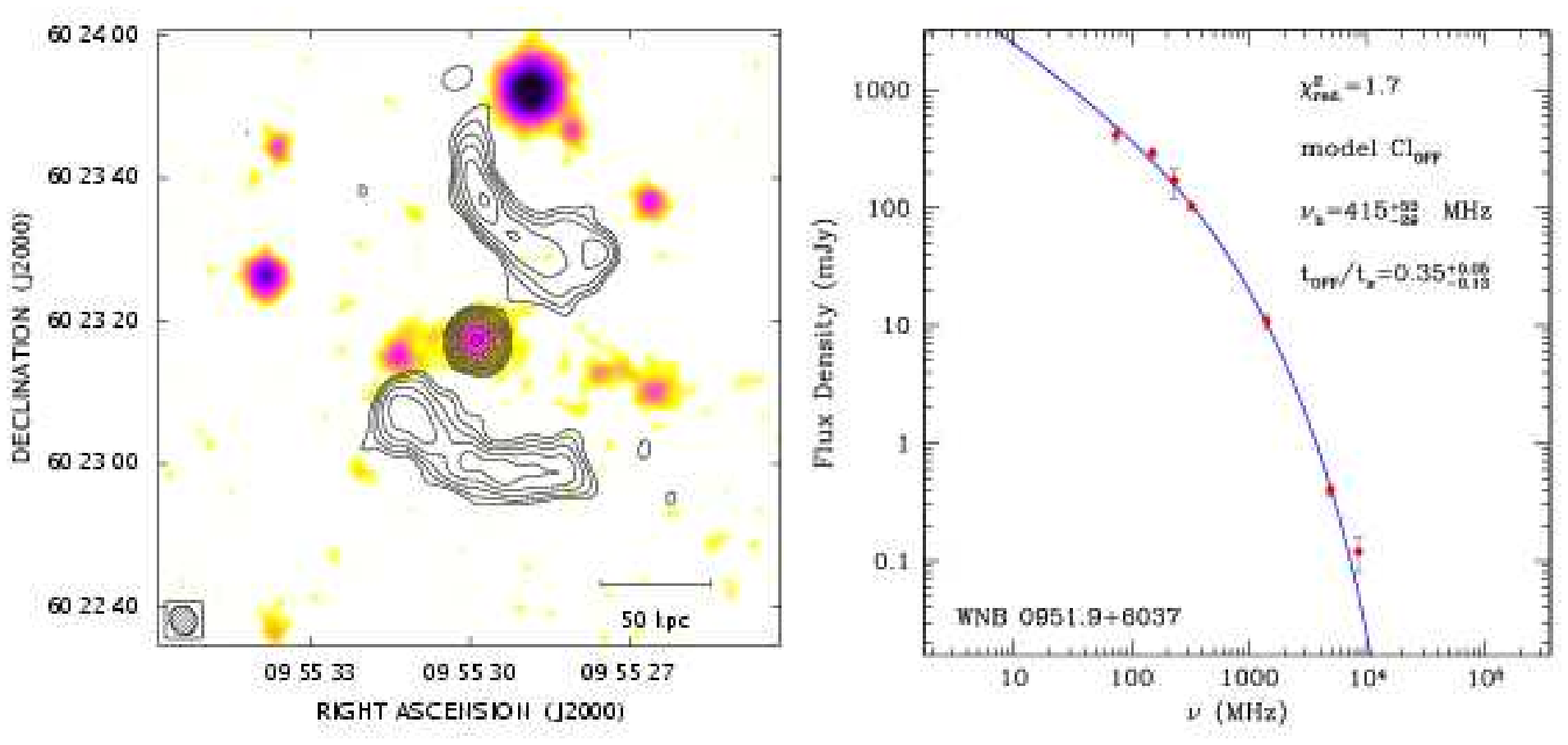}
  \caption{WNB0951.9+6037. Left: radio contours at 1.4 GHz overlayed on a DSS2 red image. Contour levels start at 
 78 $\mu$Jy/beam (3$\sigma$) and scale by $\sqrt{2}$. The beam is $4\farcs 8\times 3\farcs 6$. 
Right: integrated radio spectrum. The solid line represents the best fit of the synchrotron model described in the text.}
  \label{fig:fig2}
\end{figure*}

\begin{figure*}
  \centering
  \includegraphics[width=18cm]{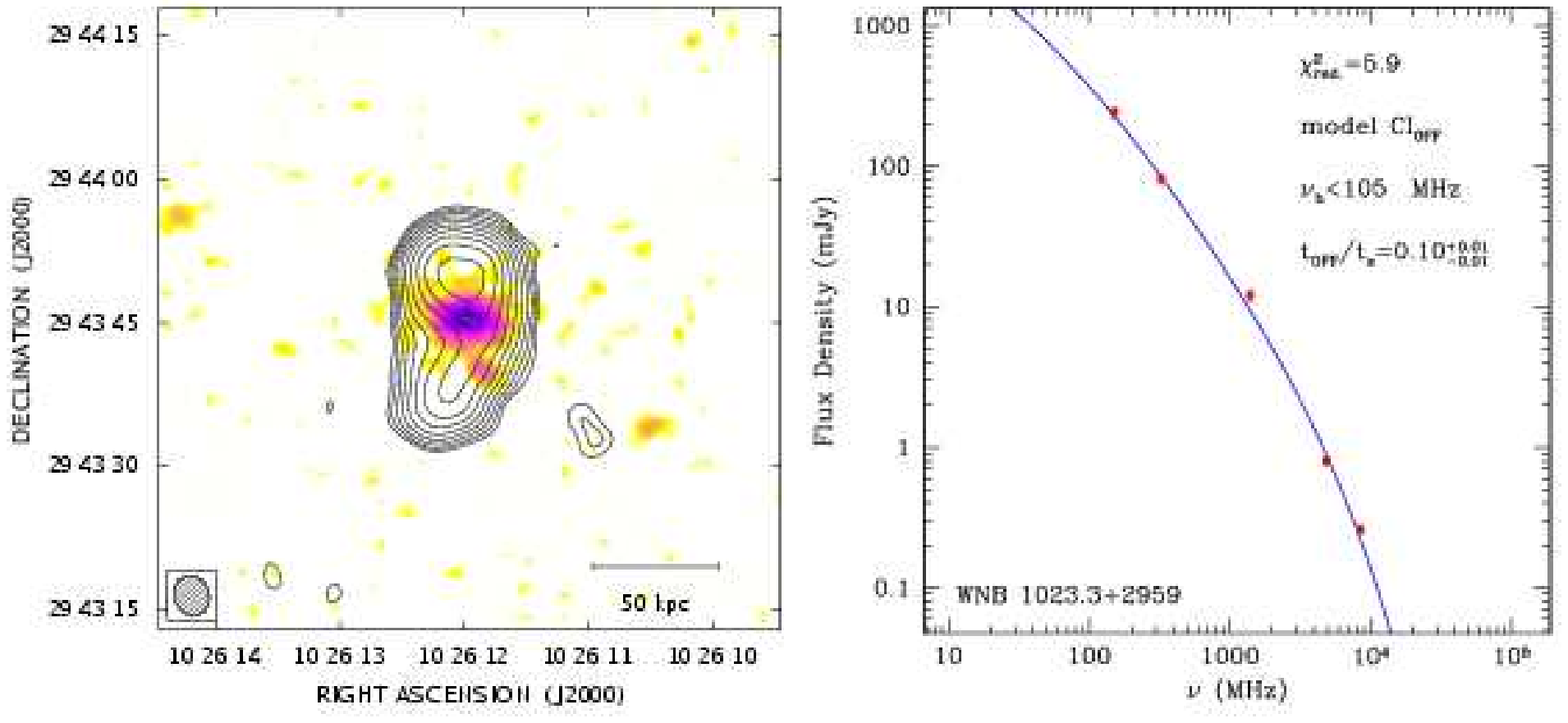}
   \caption{WNB1023.3+2959. Left: radio contours at 1.4 GHz overlayed on a DSS2 red image. Contour levels start at 
 72 $\mu$Jy/beam (3$\sigma$) and scale by $\sqrt{2}$. The beam is $4\farcs 4\times 3\farcs 8$. 
Right: integrated radio spectrum. The solid line represents the best fit of the synchrotron model described in the text.}
  \label{fig:fig3}
\end{figure*}

\begin{figure*}
  \centering
  \includegraphics[width=18cm]{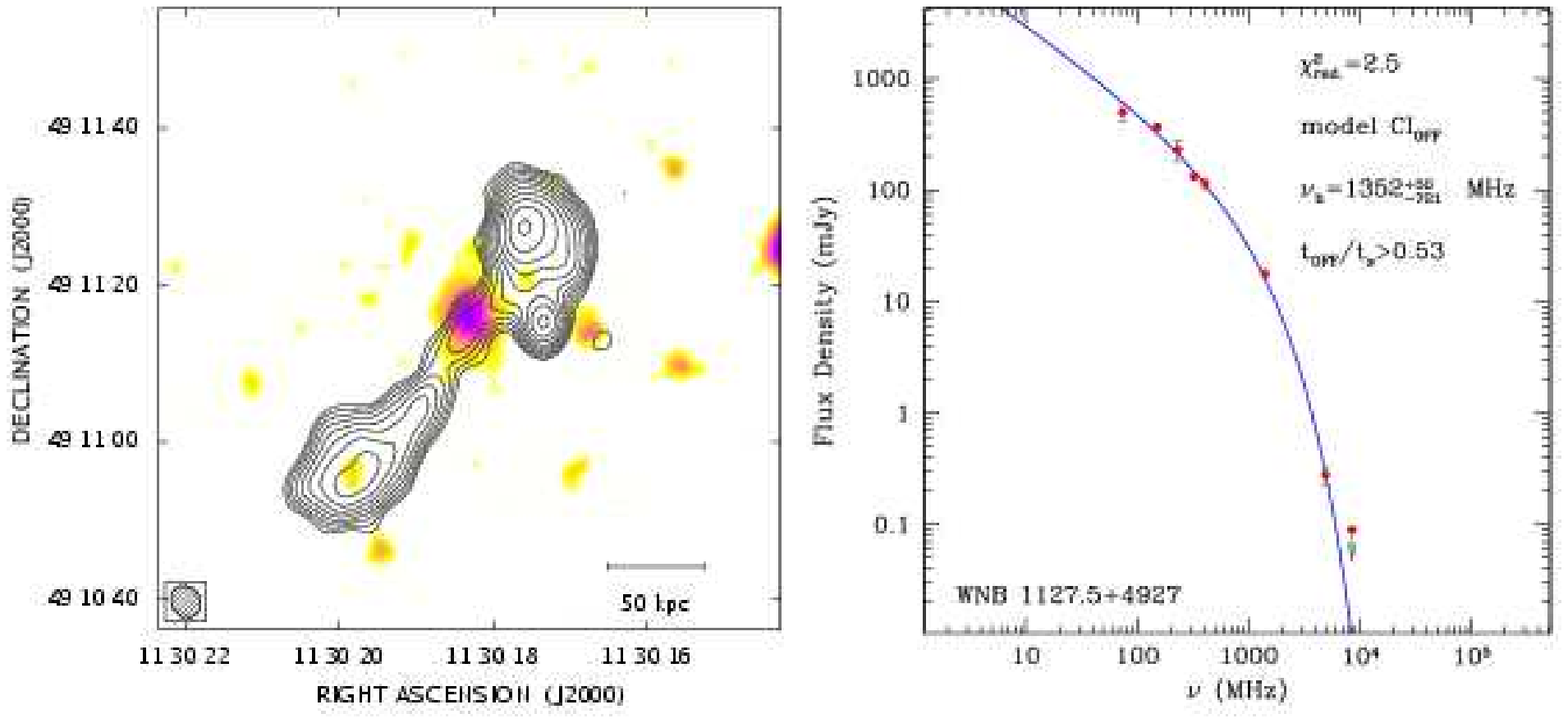}
   \caption{WNB1127.5+4927. Left: radio contours at 1.4 GHz overlayed on a DSS2 red image. Contour levels start at 
 87 $\mu$Jy/beam (3$\sigma$) and scale by $\sqrt{2}$. The beam is $4\farcs 5\times 3\farcs 7$. 
Right: integrated radio spectrum. The solid line represents the best fit of the synchrotron model described in the text.}
  \label{fig:fig4}
\end{figure*}

\begin{figure*}
  \centering
  \includegraphics[width=18cm]{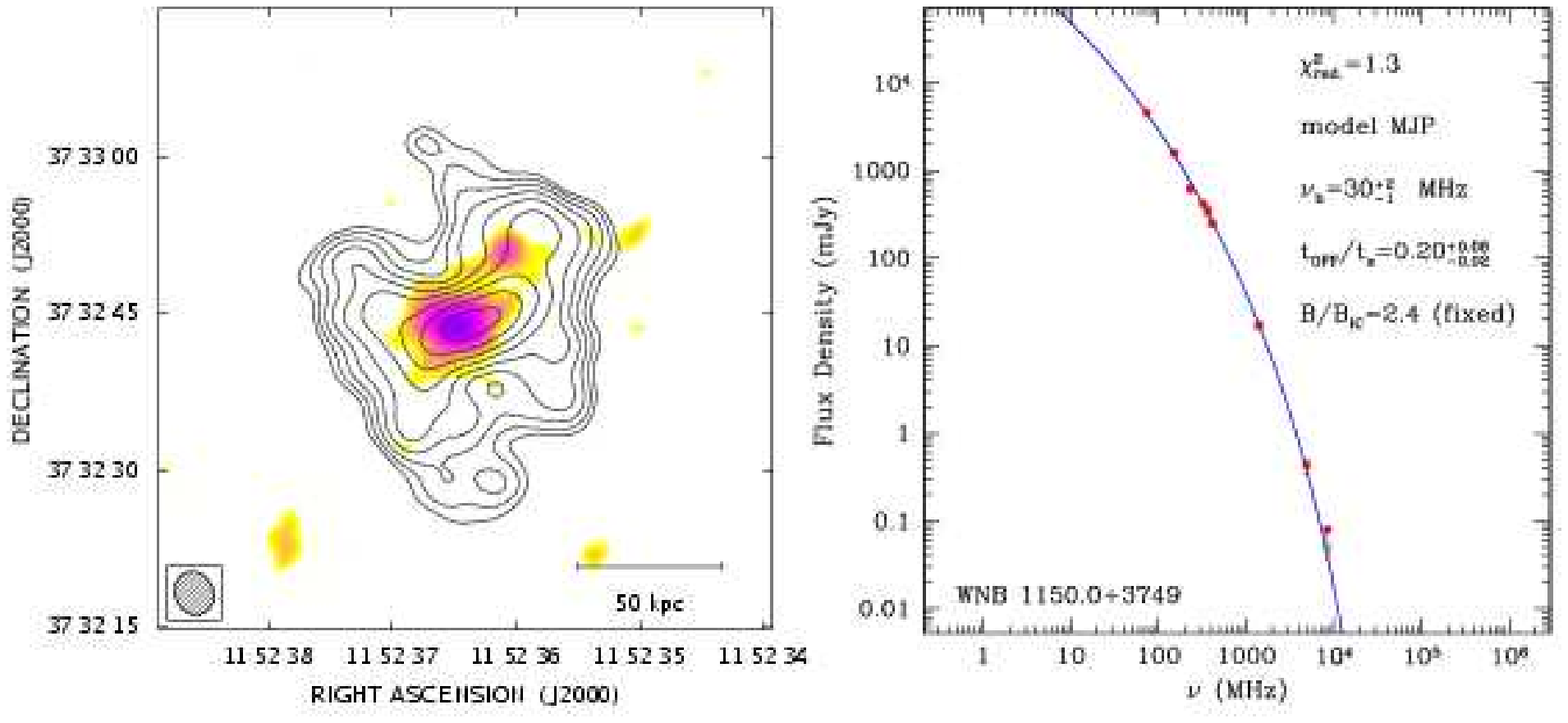}
  \caption{WNB1150.0+3749. Left: radio contours at 1.4 GHz overlayed on a DSS2 red image. Contour levels start at 
 102 $\mu$Jy/beam (3$\sigma$) and scale by $\sqrt{2}$. The beam is $4\farcs 6\times 3\farcs 9$. 
Right: integrated radio spectrum. The solid line represents the best fit of the synchrotron model described in the text.}
  \label{fig:fig5}
\end{figure*}

\begin{figure*}
  \centering
  \includegraphics[width=18cm]{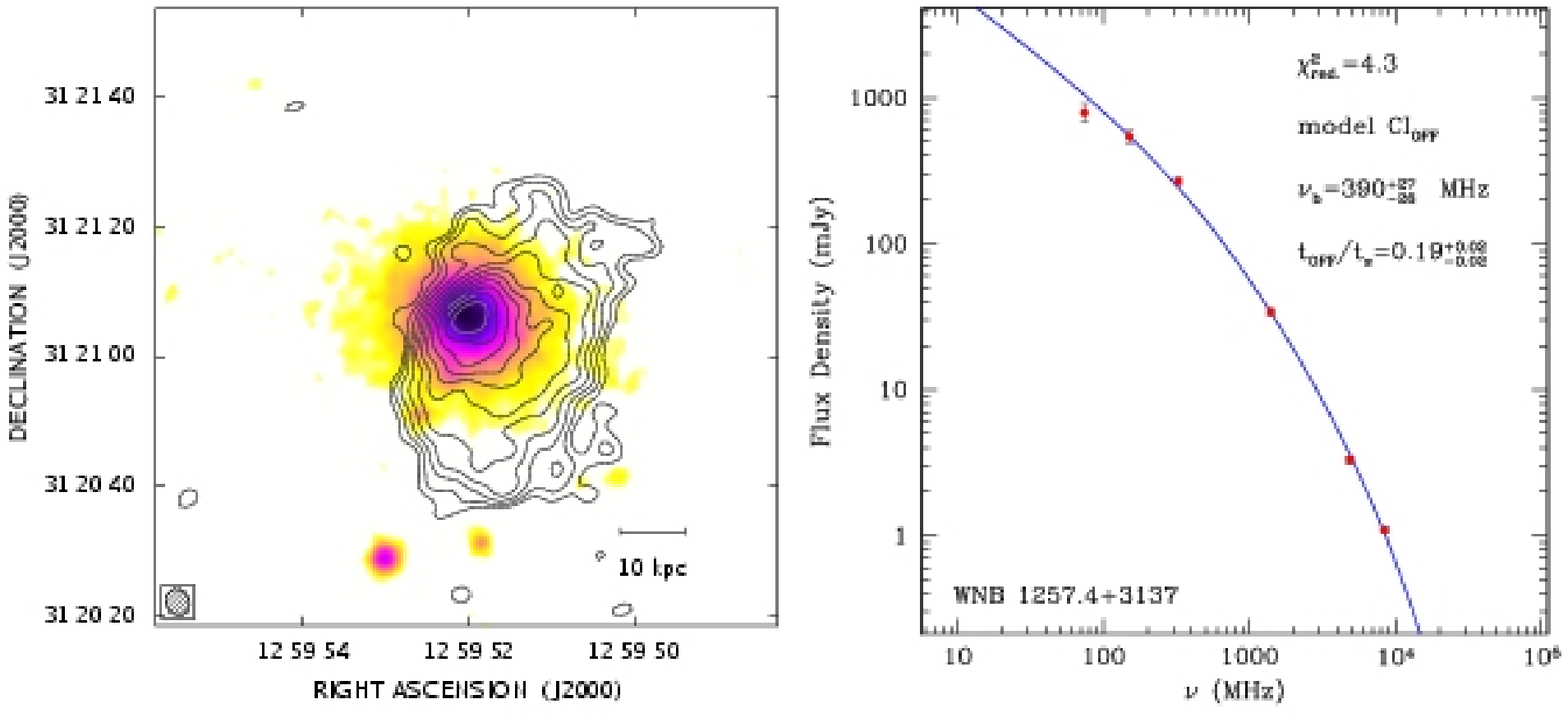}
  \caption{WNB1257.4+3137. Left: radio contours at 1.4 GHz overlayed on a DSS2 red image. Contour levels start at 
 84 $\mu$Jy/beam (3$\sigma$) and scale by $\sqrt{2}$. The beam is $4\farcs 6\times 4\farcs 1$. 
Right: integrated radio spectrum. The solid line represents the best fit of the synchrotron model described in the text.}
  \label{fig:fig6}
\end{figure*}

\begin{figure*}
  \centering
  \includegraphics[width=18cm]{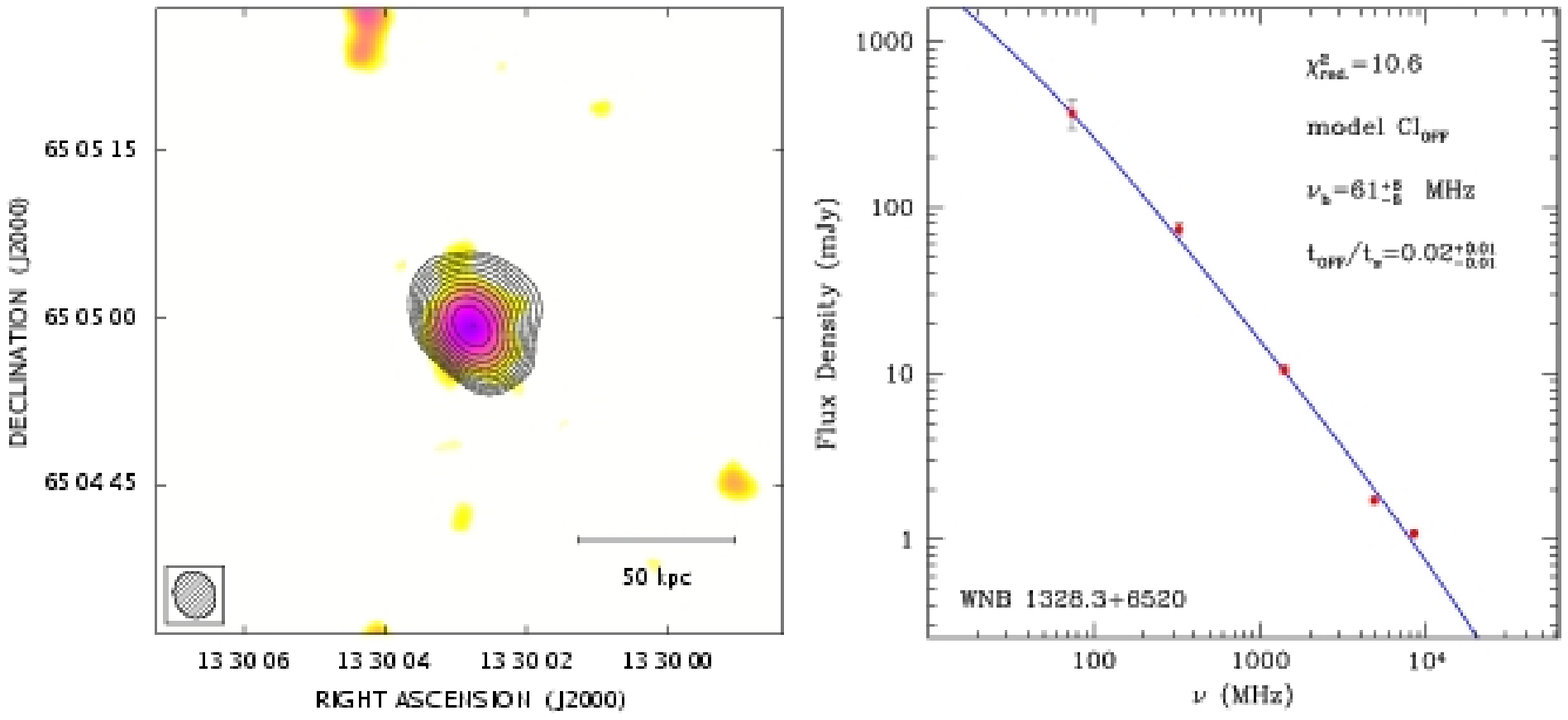}
 \caption{WNB1328.3+6520. Left: radio contours at 1.4 GHz overlayed on a DSS2 red image. Contour levels start at 
 180 $\mu$Jy/beam (3$\sigma$) and scale by $\sqrt{2}$. The beam is $5\farcs 3\times 3\farcs 6$. 
Right: integrated radio spectrum. The solid line represents the best fit of the synchrotron model described in the text.}
  \label{fig:fig7}
\end{figure*}

\begin{figure*}
  \centering
  \includegraphics[width=18cm]{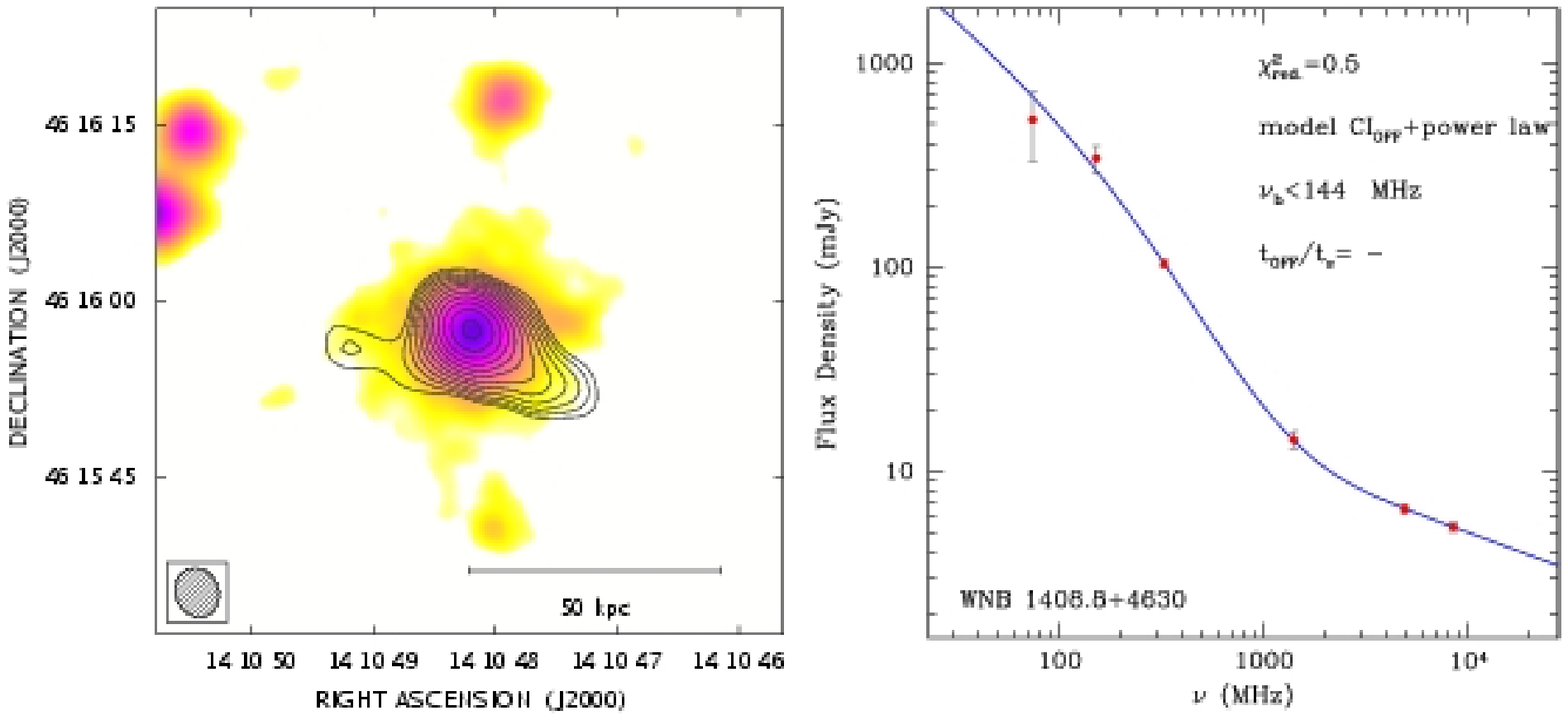}
   \caption{WNB1408.8+4630. Left: radio contours at 1.4 GHz overlayed on a DSS2 red image. Contour levels start at 
 150 $\mu$Jy/beam (3$\sigma$) and scale by $\sqrt{2}$. The beam is $4\farcs 4\times 3\farcs 7$. 
Right: integrated radio spectrum. The solid line represents the best fit of the synchrotron model described in the text.}
  \label{fig:fig8}
\end{figure*}

\begin{figure*}
  \centering
  \includegraphics[width=18cm]{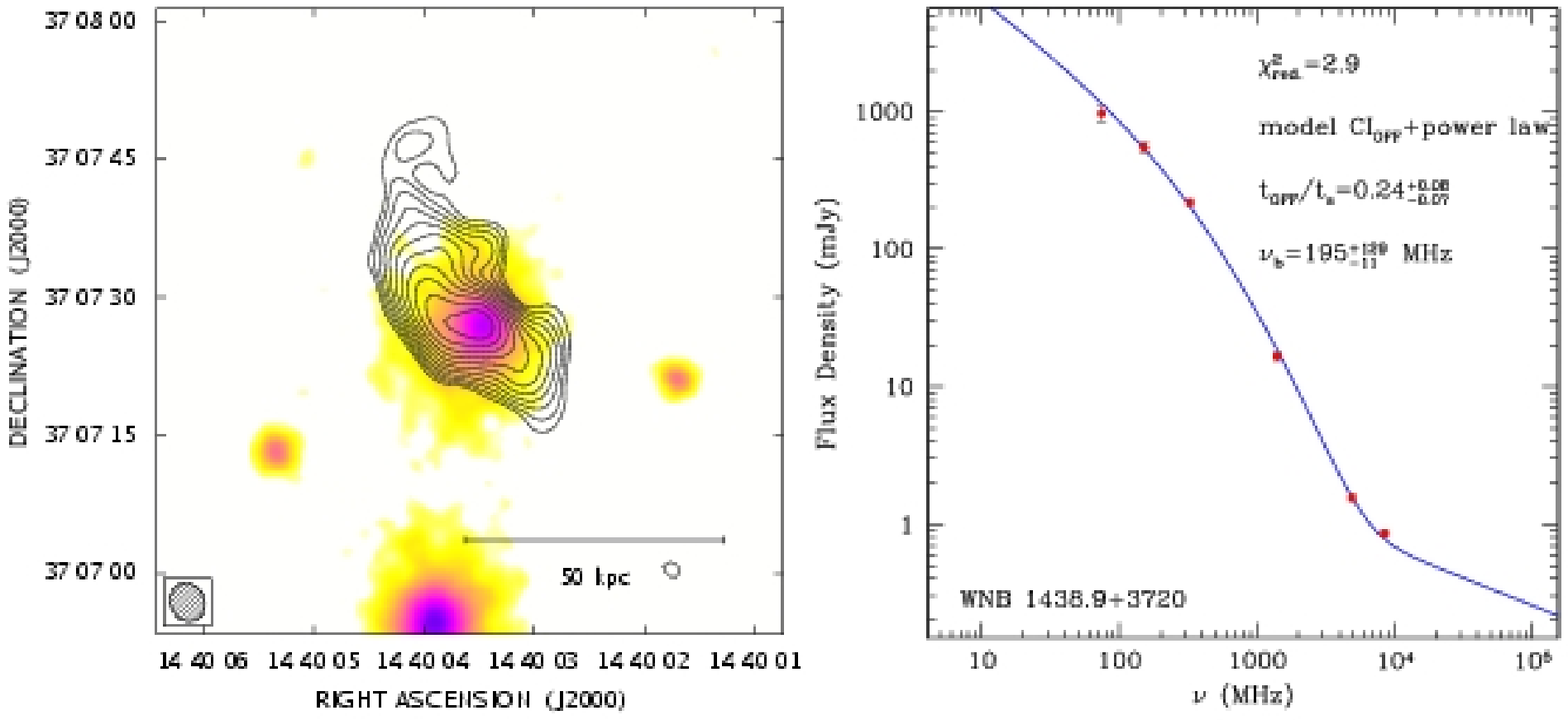}
   \caption{WNB1438.0+3720. Left: radio contours at 1.4 GHz overlayed on a DSS2 red image. Contour levels start at 
 96 $\mu$Jy/beam (3$\sigma$) and scale by $\sqrt{2}$. The beam is $4\farcs 4\times 3\farcs 7$. 
Right: integrated radio spectrum. The solid line represents the best fit of the synchrotron model described in the text.}
  \label{fig:fig9}
\end{figure*}

\begin{figure*}
  \centering
  \includegraphics[width=18cm]{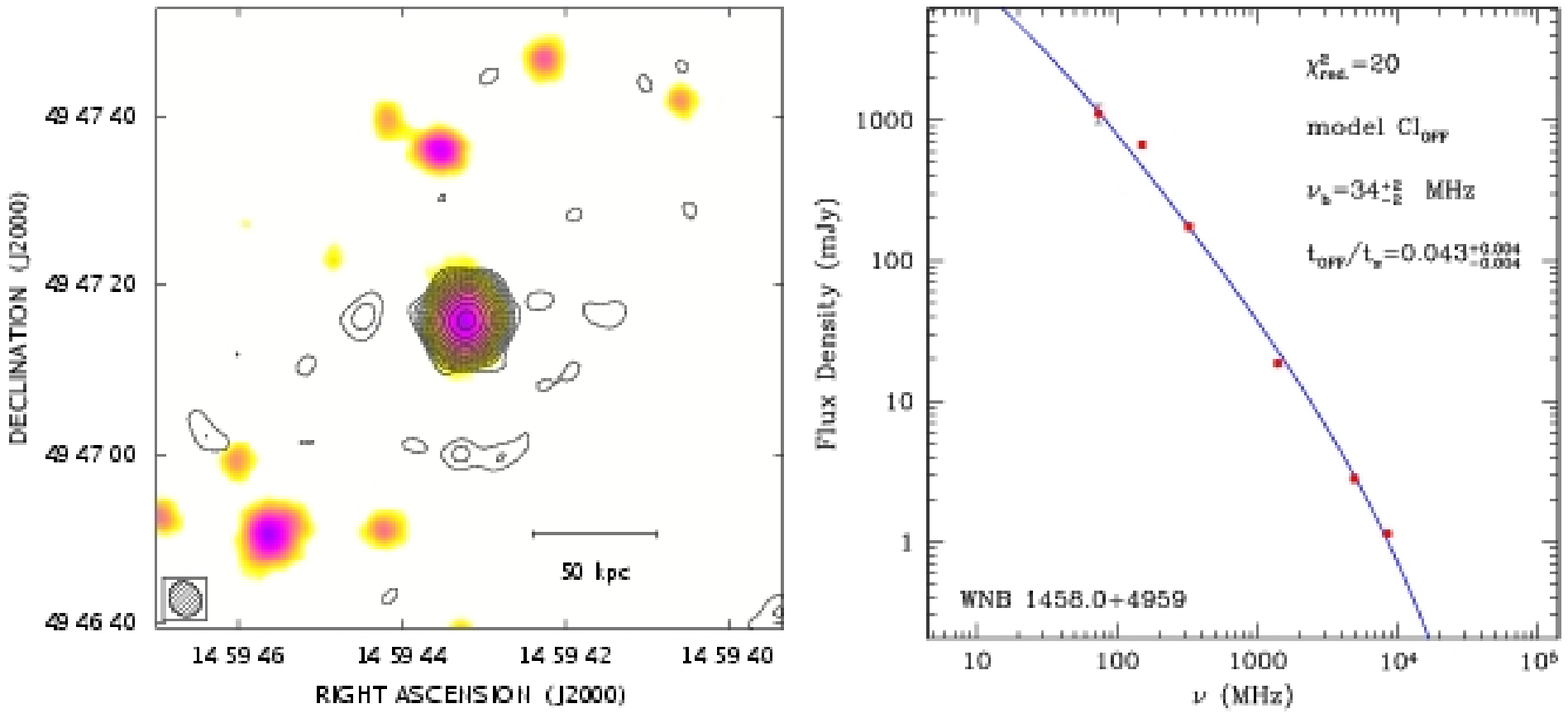}
  \caption{WNB1458.0+4959. Left: radio contours at 1.4 GHz overlayed on a DSS2 red image. Contour levels start at 
 150 $\mu$Jy/beam (3$\sigma$) and scale by $\sqrt{2}$. The beam is $4\farcs 4\times 3\farcs 7$. 
Right: integrated radio spectrum. The solid line represents the best fit of the synchrotron model described in the text.}
  \label{fig:fig10}
\end{figure*}

\begin{figure*}
  \centering
  \includegraphics[width=18cm]{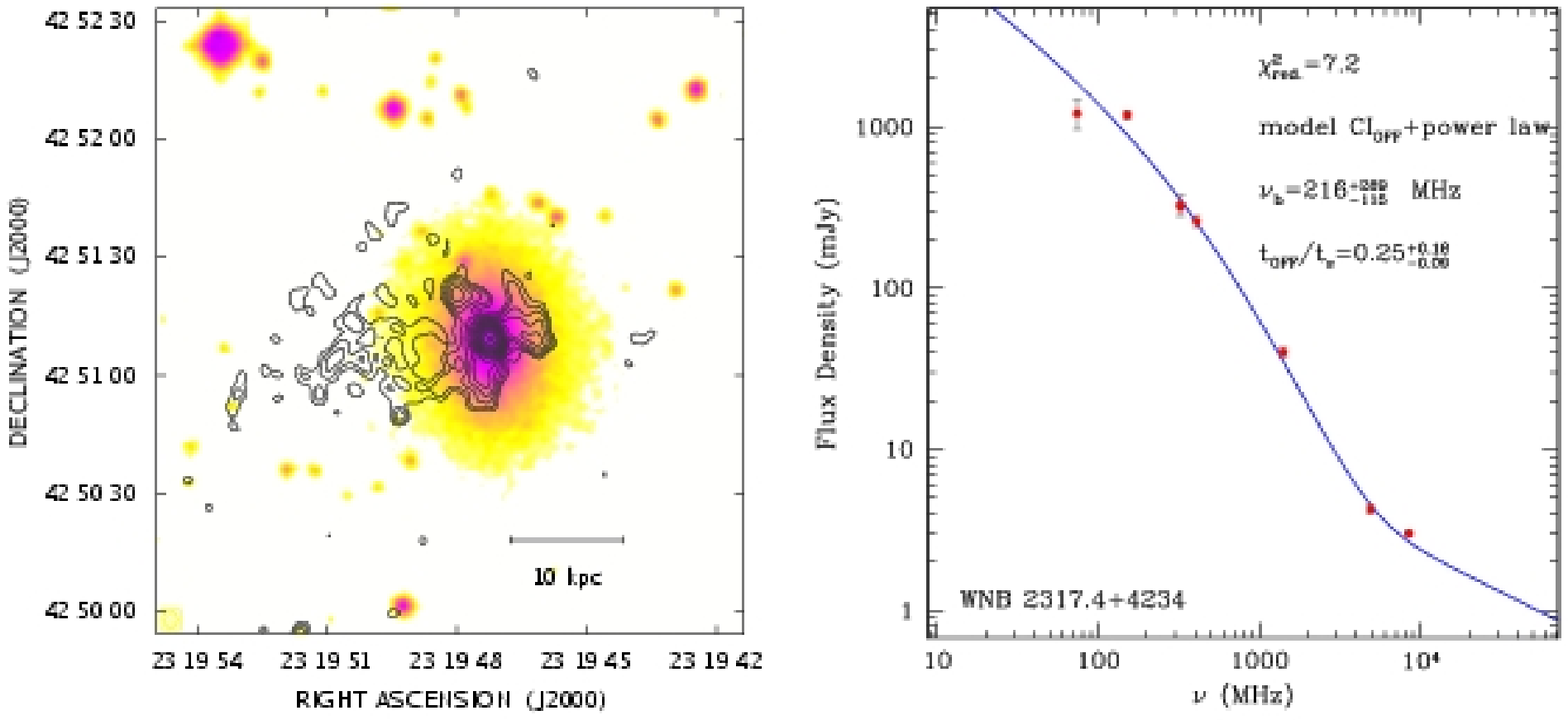}
 \caption{WNB2317.4+4234. Left: radio contours at 1.4 GHz overlayed on a DSS2 red image. Contour levels start at 
 96 $\mu$Jy/beam (3$\sigma$) and scale by $\sqrt{2}$. The beam is $4\farcs 3\times 3\farcs 7$. 
Right: integrated radio spectrum. The solid line represents the best fit of the synchrotron model described in the text.}
  \label{fig:fig11}
\end{figure*}

\subsection{ Radio observations}
\label{subs:radio}

In order to determine whether the  integrated radio spectra of these 
dying source candidates are really steep also at high frequencies we have 
observed the sample in snap-shot mode with the VLA D-array at 4.8 and 8.4 GHz. 
The sources are sufficiently compact ($ LAS < 1$ arcmin) for 
being only slightly resolved at high frequency and therefore we are able to 
measure the total flux density. However, we need to know whether active radio structures 
such as radio cores, jets and hot spots, are present. In order to fill this gap in our knowledge we
also performed VLA B-array observations, at 1.4, 4.8 and 8.4 GHz.

Four of the most promising objects were observed with the  VLA A-array at 327 MHz.
In this way we have obtained spectral information (between 1.4 GHz  
and 327 MHz) on the source components.

A summary of the observations, including the array, frequency, length of 
observation, beam size and r.m.s. noise is given in Table \ref{tab:observing log}. 
A bandwidth of 50 MHz was used for the B- and D-array, while the observations with the A-array 
were done in line mode.
 
Calibration and imaging was done using the NRAO Astronomical Image Processing System (AIPS) in the standard way.

\begin{figure*}
  \centering
  \includegraphics[width=18cm]{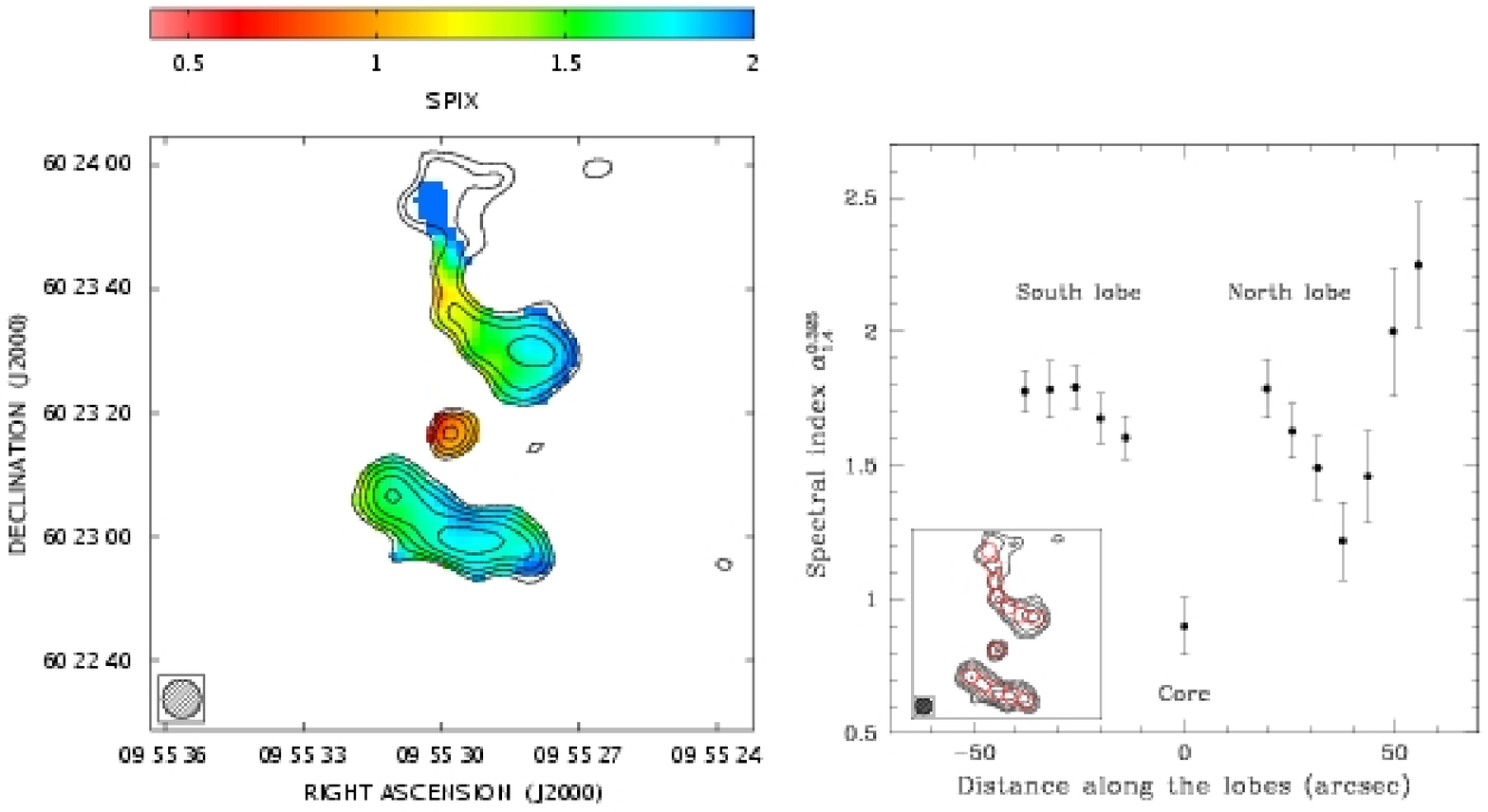}
 \caption{WNB0951.9+6037. Left panel: spectral index in grey scale, superposed on radio contours at 327 MHz. 
Radio contours start at $3\sigma$ and increase by $\sqrt{2}$.
Right panel: profile of the spectral index along the lobes (see also the text)}
  \label{fig:fig12}
\end{figure*}

\subsection{ Optical  observations}
\label{subs:optical}

Optical spectroscopy was done with the DOLORES spectrograph
installed at the Nasmyth B focus of the 3.5-m Galileo telescope on the Roque
de los Muchachos in La Palma, Spain. The observations were performed in service
mode during several nights between Jan 29 2004 and Aug 5 2005.
We used the LR-B grism and a slit size between 1.0 and 1.5 arcsec, depending
on the seeing conditions during the observations.
The dispersion was 2.8 \AA\  per pixel and the spectral resolution about 11 \AA .
The typical useful spectral range is $\sim3500$ to $\sim8000$ \AA\    with
increasing fringing beyond 7200 \AA . The slit
orientation was adjusted to maximise the number of spectra obtained with a
single setup.

\begin{figure*}
  \centering
  \includegraphics[width=18cm]{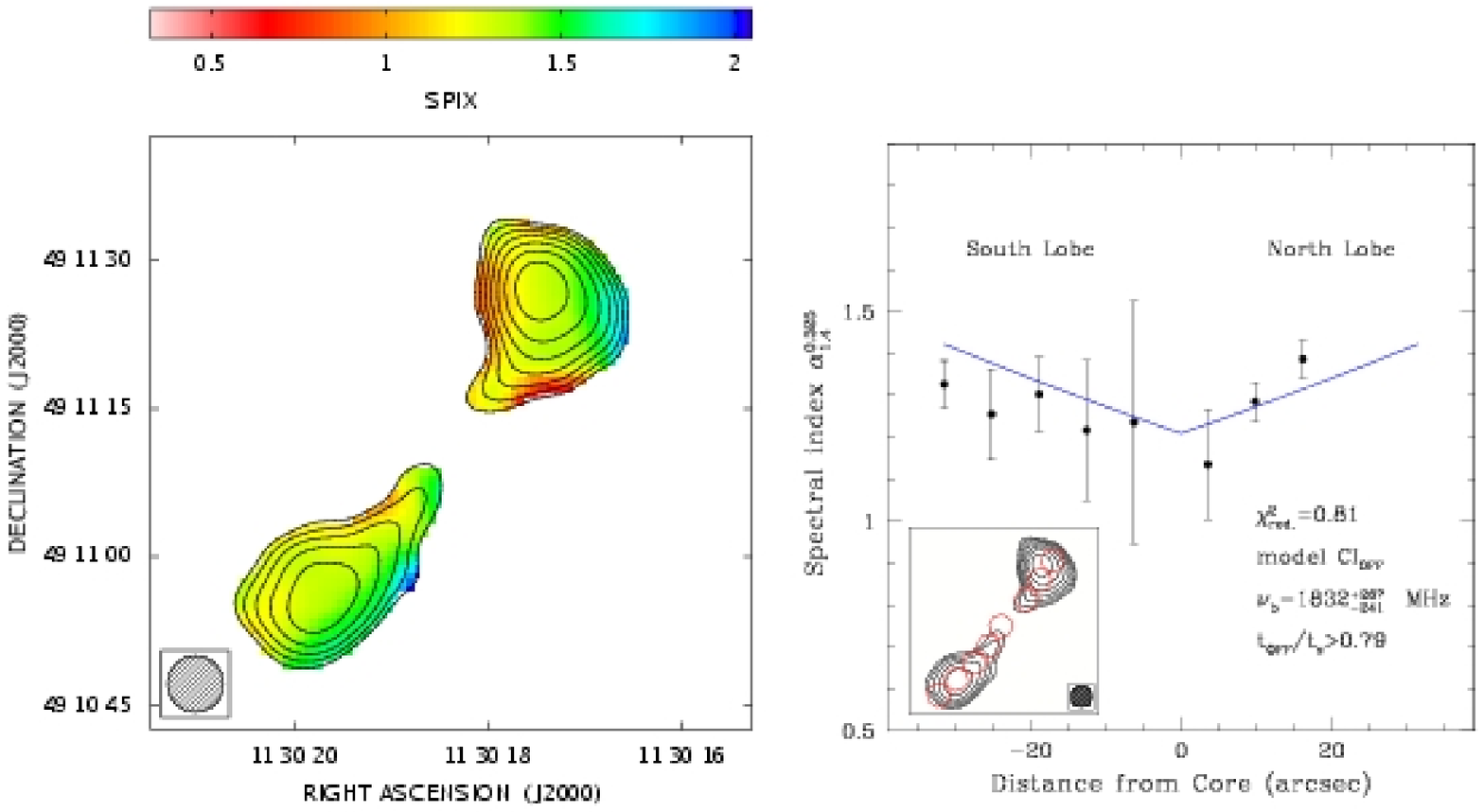}
 \caption{WNB1127.5+4927. Left panel: spectral index in grey scale, superposed on radio contours at 327 MHz. 
Radio contours start at $3\sigma$ and increase by $\sqrt{2}$.
Right panel: profile of the spectral index along the lobes (see also the text)}
  \label{fig:fig13}
\end{figure*}

The reduction of the spectra was carried out using the LONGSLIT package of 
NOAO's IRAF reduction software. A bias frame was constructed by averaging
`zero second' exposures taken at the end of the night. This was subtracted
from every non-bias frame. The pixel-to-pixel variations were calibrated 
using internal flats. 
The sky contribution was removed by subtracting a sky spectrum obtained by
fitting a polynomial to the intensities measured along the spatial direction
excluding the vicinity of objects. One-dimensional spectra were extracted by
averaging in the spatial direction over an aperture as large as the spatial
extent of the source. Wavelength calibration was carried out by measuring
the positions on the frames of known emission lines from a He-lamp and a
flux calibration was obtained from a comparison with spectra of spectro-photometric standard stars.

As the main goal of our spectroscopic observations was the determination of 
the redshifts, the exposure times were kept as short as possible. Almost all
spectra show the 4000 \AA\ break, the Ca H and K lines, as well as 
the MgI and NaI absorption lines. Most sources show no obvious emission
lines, except  WNB1408.8+463014, which may have H$\alpha$ in emission; 
however that part of the spectrum is strongly affected by fringing.
The resulting redshifts are compiled in Table  \ref{tab:observational data}.  
In addition we report the redshifts of some of the ambient objects which were 
covered by the slit (see Table \ref{tab:redshifts}).

\begin{figure*}
  \centering
  \includegraphics[width=18cm]{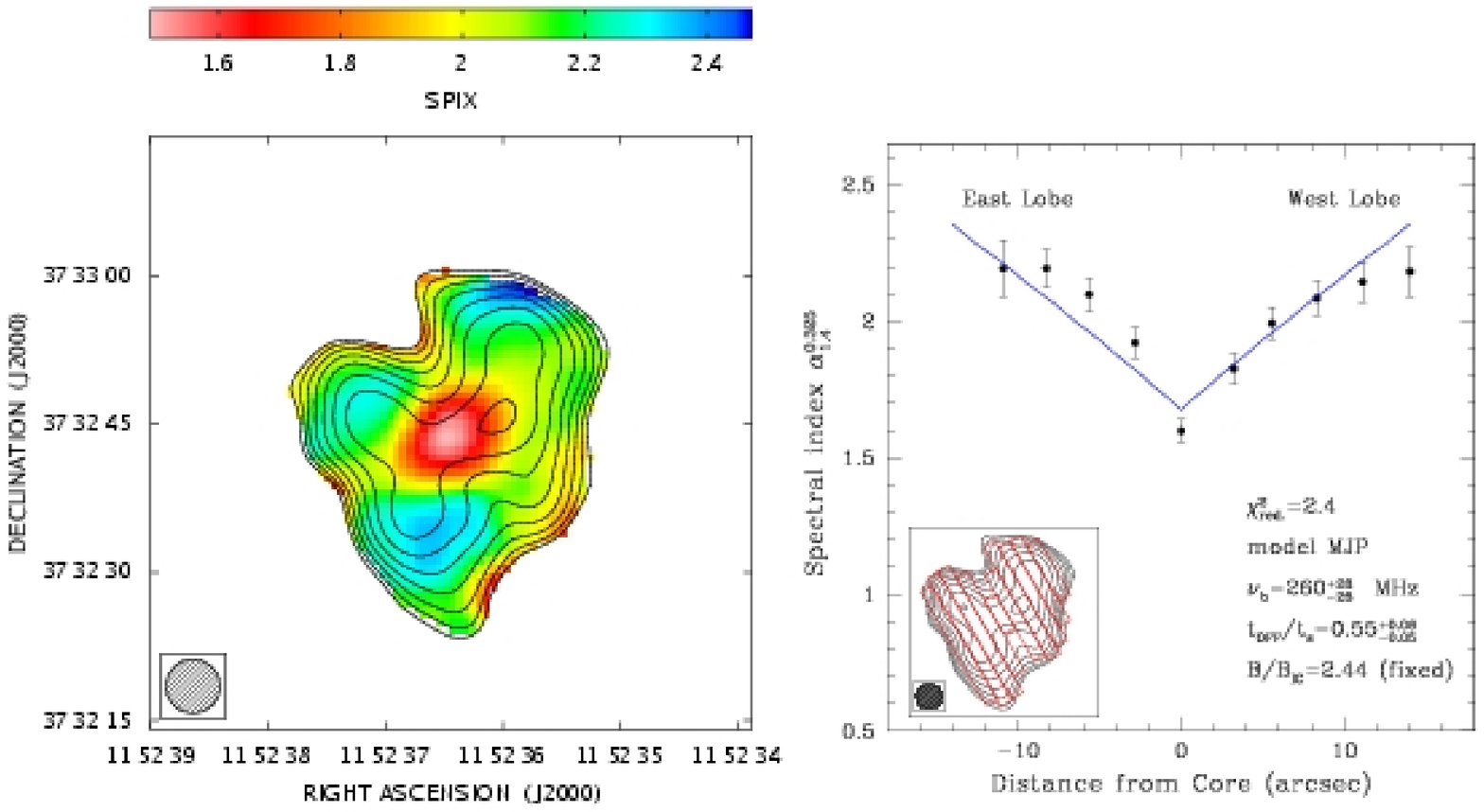}
 \caption{WNB1150.0+3749. Left panel: spectral index in grey scale, superposed on radio contours at 327 MHz. 
Radio contours start at $3\sigma$ and increase by $\sqrt{2}$.
Right panel: profile of the spectral index along the lobes (see also the text)}
  \label{fig:fig14}
\end{figure*}

\begin{figure}
  \centering
  \includegraphics[width=8cm]{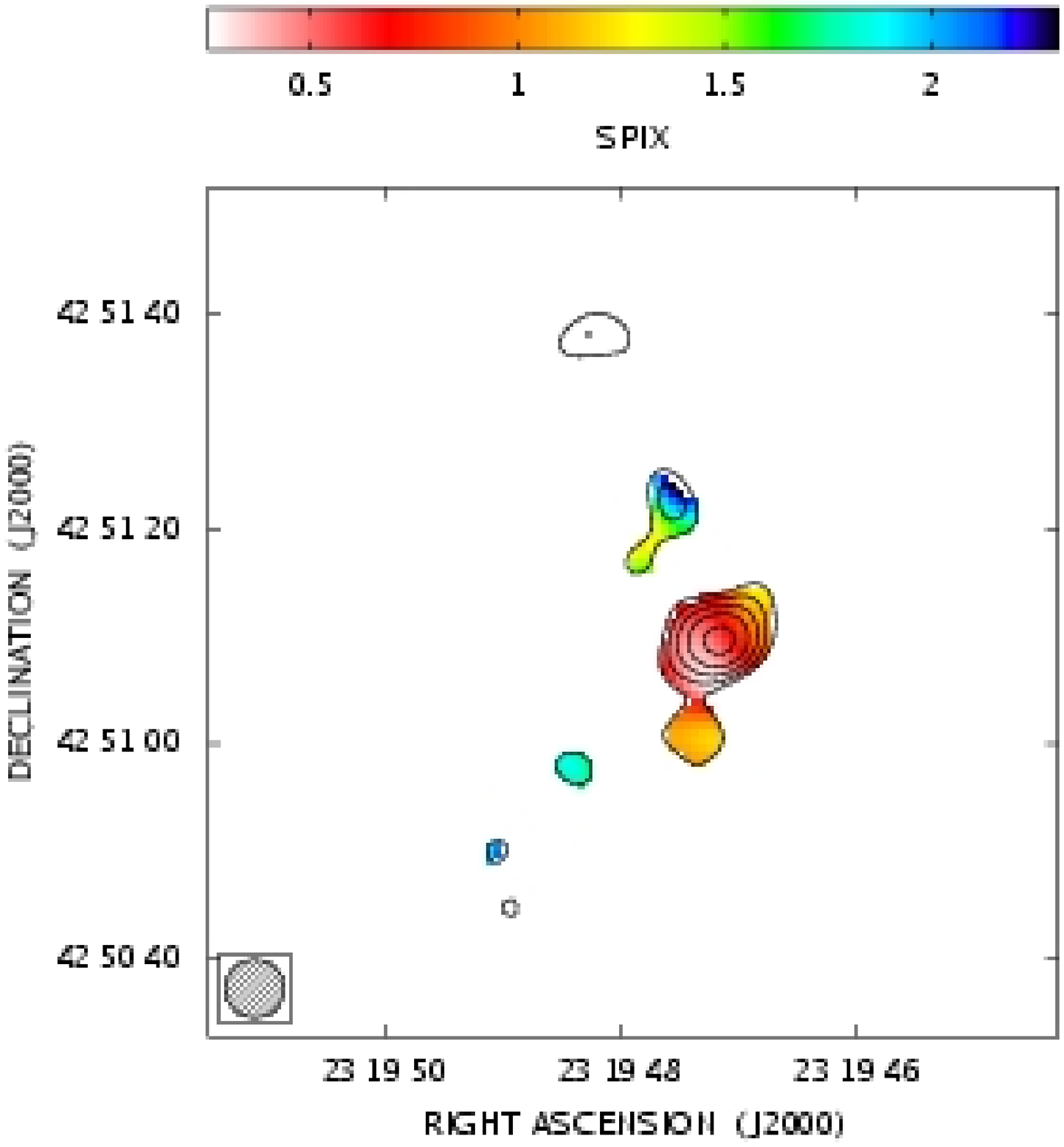}
 \caption{WNB2317.4+4234. Spectral index in grey scale, superposed on radio contours at 327 MHz.
Radio contours start at $3\sigma$ and increase by $\sqrt{2}$.
}
  \label{fig:fig15}
\end{figure}

\section{Results}
\label{sec:results}

\subsection{Radio maps and integrated spectra}
\label{subs:maps}

Images of the sources are given in Figs. \ref{fig:fig1} to \ref{fig:fig11}. 
The contours of the B-array observations at 1.4 GHz are superposed on 
DSS2 red images. 
Contour maps of the A-array observations at 327 MHz, shown in  
Figs. \ref{fig:fig12} to \ref{fig:fig15} are superposed on grey scale images
of the spectral index (between 327 MHz and 1.4 GHz).
D-array images are not shown, since all sources are unresolved.

Measured source, and if possible also component, parameters 
are given in Table \ref{tab:observational data}.
We give the position of the associated optical galaxy and of the source 
radio core when present. We also give the total flux of the source or 
that of the components together with the corresponding observing frequency.
Usually only the central component is detected  at high frequency. The size 
of the source corresponds to the full-width-half-maximum (FWHM) of a
Gaussian fit or to the largest angular size (LAS), using the lowest reliable contour.

A dying source should show exponential steepening of its total spectrum. 
The sources of our sample were selected on the basis of their spectral steepness between 
1.4 GHz and 325 MHz, but we also collected other spectral information:
in addition to the D-array observations done at 5 and 8.4 GHz, we also made use 
of CATS (the on-line Astrophysical CATalogs support System, at http://cats.sao.ru/), thus recovering data at
different frequencies. The 74 MHz flux densities were determined 
directly from NRAO VLSS images that are available on-line (at URL: http://lwa.nrl.navy.mil/VLSS/). 
The results of this search are shown in Table \ref{tab:integrated flux}, where all the available data for each 
individual source are listed. 

The integrated spectra are shown in the right-hand panels of Figs. \ref{fig:fig1} to \ref{fig:fig11}.

The total spectra of most of the radio sources in our sample show a quasi-exponential
cutoff in the frequency range between 74 MHz and 8.4 GHz. For some sources the spectral
cutoff is particularly strong (WNB0951.9+6037, WNB1127.5+4927, and WNB1150.0+3749).
In some others (WNB1328.3+6520 and WNB1458.0+4959) the curvature is less pronounced and the 
integrated spectrum is almost a power law in the considered frequency range. Finally, WNB1408.8+4630, WNB1438.0+3720 and
WNB2317.4+4234 show evident flattening of the integrated spectrum at high frequency. This flattening is due to the
presence of a core-jet component that dominates the spectrum at the highest frequencies. 
This is confirmed by the images at higher resolution (not shown), in which only the core is visible.

\subsection{Spectral indices of source components}
\label{subs:spectral index}

WNB0951.9+6037, WNB1127.5+4927, WNB1150.0+3749, and WNB2317.4+4234
were also observed at 327 MHz with the VLA A-array.
Spectral index maps, using both 327 MHz and 1.4 GHz data at 6 arcsec resolution, are shown in the left-hand
panels of Figs. \ref{fig:fig12} to \ref{fig:fig15}. In the right-hand panels the spectral index distributions are shown, as
computed in boxes along the ridge line, or all over the source, in the way shown in the inset.

In WNB0951.9+6037 the northern lobe (or tail) has $\alpha _{0.33}^{1.4} = 1.2$ 
in the centre, and steepens going outwards in both directions to values of   
$\alpha _{0.33}^{1.4} = 1.9$ and $\alpha _{0.33}^{1.4} = 2.2$ at the respective ends.
The south lobe has a uniform spectral index of $\alpha _{0.33}^{1.4} \sim 1.7$.
Note that also the central component of this source has a steep spectrum, with $\alpha \sim 0.9$.
In WNB1127.5+4927 the spectral indices of the two lobes are almost constant with values of the order of 1.3. At most
there might be a very slight steepening of no more than $\sim 0.2$ (see Fig. \ref{fig:fig13}).
In WNB1150.0+3749 the spectral index varies from  $\alpha _{0.33}^{1.4} = 1.6$ in the source centre to  
$\alpha _{0.33}^{1.4} = 2.2$ in the outer parts.
In WNB2317.4+4234 only the head of the tail is visible at 327 MHz; it has $\alpha _{0.33}^{1.4} = 0.6$.

\section{Comments on individual sources}
\label{sec:comments}
\begin{itemize}

\item WNB0722.4+4129:  a small asymmetric double without a core. It is completely resolved out 
in the B-array observations at 5 and 8.4 GHz. 

\item WNB0951.9+6037: the central 
component is detected in the B-array observation at 8.4 GHz but 
not at 4.8 GHz due to the higher rms noise on the 4.8 GHz map. Since it has $\alpha_{0.33}^{1.4} = 0.9 $ 
and $ \alpha_{1.4}^{8.4} = 1.2$ it is difficult to consider this an active core.
At 327 MHz and 1.4 GHz the source structure is very similar, the northern component being slightly 
more elongated at 327 MHz. 

\item WNB1023.3+2959:  a small asymmetric double without  core. It is completely resolved out in the 
B-array observations at 5 and 8.4 GHz.  

\item WNB1127.5+4927: a classical double source with a compact (apparently background) 
source superimposed on the western lobe. No core has been detected. 
At 5 and 8.4 GHz (B-array) only the background source is detected.  The double source structure is very 
similar at 327 MHz and 1.4 GHz, but the background source is not detected at 327 MHz due to its flat spectral index ($\approx 0.2$).

\item WNB1150.0+3749:  a source with amorphous structure. It may 
be a small double source without a core and is completely resolved out in 
the B-array observations at 5 and 8.4 GHz. At 327 MHz the radio 
structure is very similar to the one at 1.4 GHz. 

\item WNB1257.4+3137: a source with amorphous structure at 1.4 GHz. 
At 4.8 GHz only the central component is still visible, which, however, is completely 
resolved out at 8.4 GHz. 

\item WNB1328.3+6520: a compact radio source with a very steep spectrum. It is 
detected in all B-array observations. This source resembles a Compact Steep Spectrum (CSS)
source, but with an unusually steep spectrum. 

\item WNB1408.8+4630: apparently this source has one-sided morphology. 
The galaxy position coincides with the bright compact component. Much fainter 
emission extends to the south-west. Perhaps this is a core-jet radio source or,
alternatively, a small head-tail source. The bright component has $\alpha_{5.0}^{8.4} = 0.2$ 
and is probably the radio core. The very steep spectrum low frequency emission probably comes from
an extended component that is not seen at frequencies $>1$ GHz.
However, the integrated spectrum shows flattening at high frequencies due to
the presence of the flat-spectrum core. 

\item WNB1438.0+372: a small bent radio source at 1.4 GHz. A core ($\alpha_{5.0}^{8.4} = 0.4$) is detected at higher frequencies.
This radio source may be a small wide angle tail (WAT).

\item WNB1458.0+4959: a very compact radio source at 1.4 and 5.0 GHz. It is not detected at 8.4 GHz.
This source resembles a CSS source, but with an unusually steep spectrum. 

\item WNB2317.4+4234: a head-tail source. The tail is very weak and only the head is detected at 5 and 8.4 GHz. 
At 327 MHz the tail cannot be seen because the rms noise of the map is rather high. 
The head has $\alpha_{0.33}^{1.4} = 0.7$ and $\alpha_{5.0}^{8.4} =0.5$.
The associated galaxy is NGC\,7618.
\end{itemize} 

\begin{figure*}
  \centering
  \includegraphics[width=16cm]{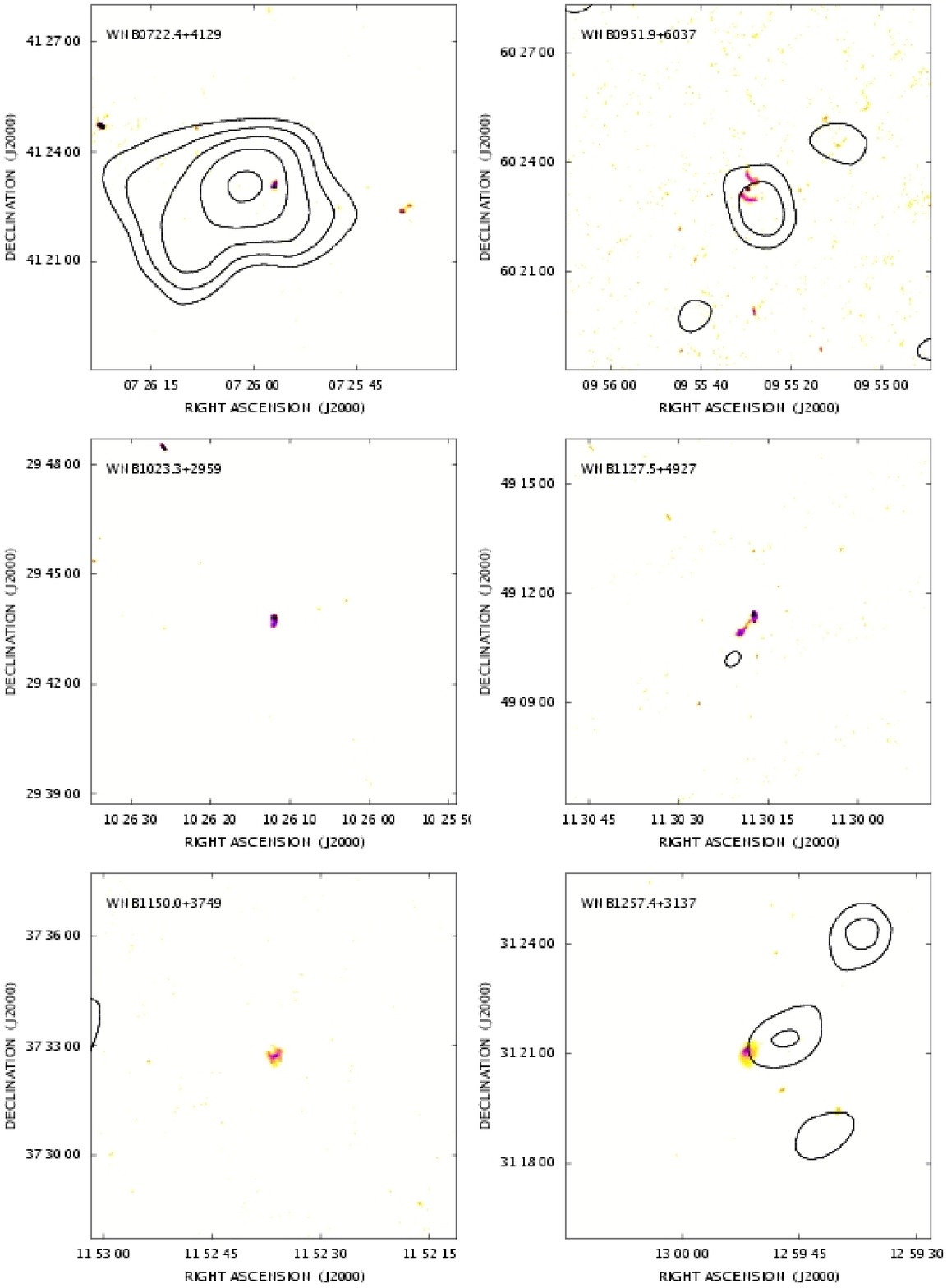}
 \caption{The VLA 1.4\,GHz B-array image (grey scale), superposed on the RASS X-ray contours. 
Contour levels start from $10^{-3}$ cts/s ($3\sigma$) and scale by $\sqrt{2}$.}
  \label{fig:fig16}
\end{figure*}

\begin{figure*}
  \centering
  \includegraphics[width=16cm]{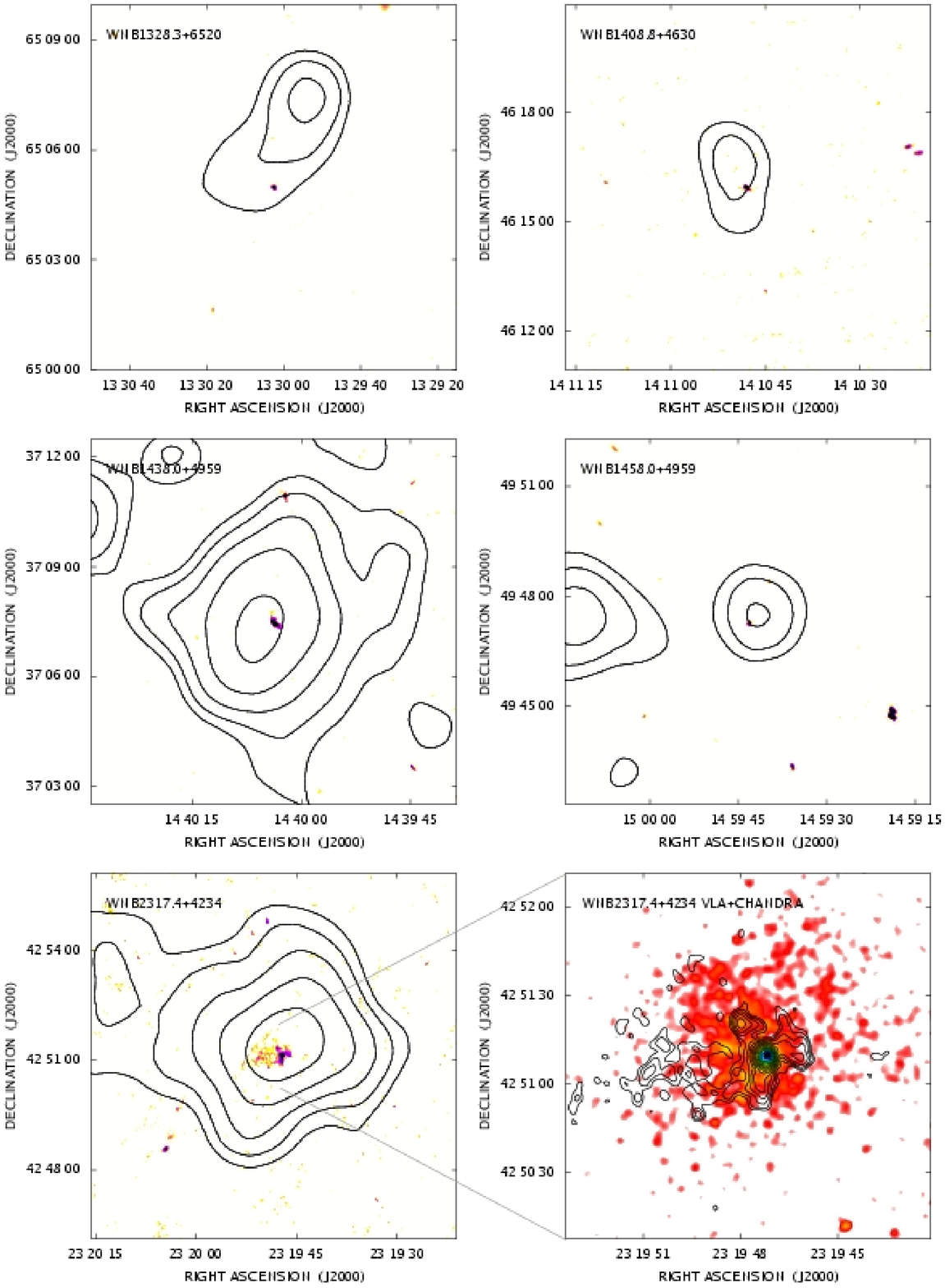}
 \caption{First five panels: the VLA 1.4\,GHz B-array image (grey scale), superposed on the RASS X-ray contours. 
Contour levels start from $10^{-3}$ cts/s ($3\sigma$) and scale by $\sqrt{2}$. Panel at bottom-right: overlay of the Chandra
image and radio contours at 1.4 GHz}
  \label{fig:fig17}
\end{figure*}

\section{Discussion}
\label{sec:discussion}

\subsection{Morphological properties and radio cores}
\label{sec:morph}

>From a morphological point of view the sources discussed here form a rather mixed bag.
The only common characteristic is their small angular size, but of course this is due to the selection
bias ($< 1$ arcmin). Since the sources cover a relatively large redshift range (from 0.02 up to about 0.26) the
linear sizes show more variety, although in a power-linear size plot these sources mix with, for example, 
the general sample of B2 radio galaxies.
It should be noted however, that the morphologies found in this sample are very different from those of the
B2 radio galaxies, but, again, this may be due to the size selection imposed on the sample of candidate dying sources.

A flat spectrum radio core is unambiguously present only in a minority of cases:
using the higher frequency data we detected a radio core in three sources, while we can put upper limits on the core 
luminosity for the other ones. Radio cores might still be present at lower levels without flattening the spectrum significantly. 
We find that nine sources have a ratio $S_{core}^{4.8 GHz}/S_{tot}^{325 MHz} \le 7 \times 10^{-3}$ (of which one is a detection at the level of 
$\sim 5 \times 10^{-3}$), while only two are above this (of order $10^{-2}$).
>From the relation between core and total luminosity given by Giovannini et al. (\cite{Giov88}) we would have expected 
a ratio $\ge 7 \times 10^{-3}$ for about half of the sources. Therefore, even allowing for the selection bias introduced by the
steepness of the spectra, the cores in the present sample are quite weak and this fact strengthens the suspicion that the 
majority in our sample may be genuine dying radio sources.  

\subsection{Equipartition parameters}
\label{equip}

We assume that the radio sources contain relativistic particles and magnetic fields uniformly distributed and in energy equipartition 
conditions. The equipartition parameters (total energy $U_{eq}$, energy density $u_{eq}$ and magnetic field $B_{eq}$) are generally 
computed assuming that the relativistic particle energies are confined between a minimum $E_{min}$ and a maximum $E_{max}$,
corresponding to the observable radio frequency range, typically 10 MHz - 100 GHz (see, e.g., Pacholczyk \cite{Pach70}). 
This choice minimizes the source energetics required by the observed radiation in the radio band. However, a fixed frequency range 
corresponds to an energy range that depends on the source magnetic field, which may change from source to source. 
Furthermore, as pointed out by Brunetti et al. (\cite{Brunetti97}) a {\it fixed frequency range} computation would miss the 
contribution from lower energy electrons, since the $E_{min}$ corresponding to 10 MHz is larger than 200 MeV for $B_{eq} \le 30~\mu G$; this
missing contribution can be very large. Because of this we have computed, wherever possible, the equipartition parameters 
assuming a fixed low energy cutoff $E_{min}$ = 10 MeV (for this choice see Brunetti et al. \cite{Brunetti97}). Intrinsic parameters, including 
equipartition parameters, are listed in Table \ref{tab:intrinsic parameters}.

For a power law energy distribution $N(E) \propto E^{-\delta}$ ($\delta = 2 \alpha + 1$) the equipartition magnetic field
is given by the following expression

\begin{equation}
B_{eq} = \left\{ F(\alpha) (1+k) \frac{L(\nu^*) (\nu^*)^{\alpha}}{V}\left(\frac{E_{min}^{1-2\alpha} -
E_{max}^{1-2\alpha}}{2 \alpha -1}\right)\right\}^{1/(3+\alpha)}
\label{beq}
\end{equation}
\smallskip
\noindent
where $\alpha$ is the spectral index (assumed to be 0.8),
$V$ the source volume in kpc$^3$, $\nu^*$ a reference frequency in GHz, $L(\nu^*)$ the monochromatic luminosity 
at $\nu^*$ in W/Hz, $k$ the ratio of proton to electron energy  (assumed to be one) and $F(\alpha)$ a  quantity in the range  
$ \approx 10^{-39} ~-~ 10^{-38}$ for $0.5 \le \alpha \le 1.0$. 

As far as the choice of $E_{max}$ is concerned (see Sect. \ref{modspectra} for details about the parameters used) we note that beyond $E(\nu_{b2})$ no 
electrons are present and the energy contribution in the range from  $E(\nu_{b})$ to $E_{max}$ is always much less than 
that below $E(\nu_{b})$. Furthermore it turns out that in the equipartition formula $E_{max} \approx E(\nu_{b})$ can be neglected.

The equipartition parameters computed in this way are reported in Table \ref{tab:intrinsic parameters}. A comparison with fixed frequency range 
calculations shows that the values of our $B_{eq}$ are larger by up to a factor two.

\subsection{Modelling the integrated radio spectra}
\label{modspectra}


The radiative energy losses are assumed to be dominant with respect to other processes (e.g. adiabatic losses).
The pitch angles of the radiating electrons are assumed to be continually isotropized in a time that is shorter 
than the radiative time scale.
According to this assumption the synchrotron energy losses are statistically the same for all electrons.
After its birth the source is supposed to be fuelled at a constant rate (i.e. {\it the continuous injection phase}) by the nuclear activity, for a duration $t_{ci}$. 
The injected  particles are assumed to have a power law energy spectrum $N(E) \propto E^{-\delta_{inj}}$, which will result in a power law radiation spectrum 
with spectral index  $\alpha_{inj} = (\delta_{inj} - 1)/2$.
In this phase the source radio spectrum changes as a function of time in a way described by the shift of break frequency 
$\nu_{b}$ to ever lower values as the time, $t_s$, increases:

\begin{equation}
\nu_{b}\propto \frac{B}{(B^2+B_{\rm IC}^2)^2 t_{s}^2}
\label{vb}
\end{equation} 
where $B$ and $B_{IC}$ are the source magnetic field and the inverse Compton
equivalent magnetic field, respectively. Below and above $\nu_{b}$ the spectral indices are
respectively $\alpha_{inj}$ and $\alpha_{inj}$+0.5.

At the time $t_{ci}$ the power supply from the nucleus is switched-off. 
After that a new phase of duration $t_{\rm OFF}$ begins (i.e. the {\it  relic phase}). A new break 
frequency $\nu_{b2}$ then appears, beyond which the radiation
spectrum drops exponentially. This second high frequency break is
 related to the first by:

\begin{equation}
\nu_{b2}=\nu_{b} \left(\frac{t_{s}}{t_{\rm OFF}}\right)^{2}
\label{vb1vb2}
\end{equation}
 
where $t_{s}=t_{ci}+t_{\rm OFF}$ is the total source age (see e.g. Komissarov \&
 Gubanov \cite{Kom94}; Slee et al. \cite{Slee01}).

Thus, the above synchrotron model (hereafter CI$_{\rm OFF}$) is described by four parameters:
\begin{itemize}
\item[i)] $\alpha_{inj}$, the injection spectral index;
\item[ii)] $\nu_{b}$, the lowest break frequency;
\item[iii)] $t_{\rm OFF}/t_{s}$, the relic to total source age ratio;
\item[iv)] $norm$, the flux normalization.
\end{itemize}

In the CI$_{\rm OFF}$ model the magnetic field strength is assumed to be uniform within the source. 
Slee et al. (\cite{Slee01}) and Murgia (in prep.) propose a more sophisticated
model (MJP), which takes into account diffusion of synchrotron electrons in a magnetic field with a Gaussian distribution of 
field strengths.
Compared to the basic CI$_{\rm OFF}$ model the MJP model is characterized by two more free parameters:

\begin{itemize}
\item[v)] $B_{\rm rms}/B_{\rm IC}$, the ratio of the rms to the inverse Compton magnetic field;
\item[vi)] $D_{\rm eff}$, the electron diffusion efficiency.
\end{itemize}

If the rms magnetic field is stronger than the inverse Compton magnetic field ($B_{\rm rms} > 5 B_{\rm IC}$) and the 
electron diffusion is low ($D_{\rm eff}\simeq 0$) the spectrum beyond $\nu_{b2}$ is less steep than the exponential
drop in the CI$_{\rm OFF}$ model. The differences in the shape of the spectral cutoff are significant only for extremely
steep spectrum sources, i.e. if $\nu_{b}< 100$ MHz and $t_{\rm OFF}/t_{s} > 0.2$. For this reason we fitted
the total spectra of all sources with the CI$_{\rm OFF}$ model, excluding WNB1150.0+3749. 
In this case only the fit obtained by the MJP model was definitely better. This is not so surprising as WNB1150.0+3749 is the source
with the steepest spectrum of all.
The model fit is represented by a solid line in the
right-hand panels of Figs.\,1 to 11, where we also show the reduced $\chi^{2}$ and the best fit parameters.
The most difficult parameter to constrain is $\alpha_{inj}$, since its determination requires flux density measurements at 
frequencies below 74 MHz, our lowest frequency. For this reason, we preferred to keep its value fixed to
 $\alpha_{inj}=0.7$ in the fitting procedure. This value is characteristic of the spectra of jets and hot spots of many active radio 
galaxies and can be assumed as a plausible injection spectral index. If we used $\alpha_{inj}=0.5$ instead, we  obtain similar 
results, although the average $\chi^{2}$ is then slightly worse. 
Beyond the high-frequency spectral cutoff in WNB1150.0+3749 the spectral drop is flatter than exponential. Therefore, this spectrum 
was fitted with an MJP model in which we fixed the diffusion efficiency at $D_{\rm eff}=0$ and the ratio at $B_{\rm rms}/B_{\rm IC}=2.4$. 
The value of $B_{\rm rms}/B_{\rm IC}$ was derived by adopting for $B_{\rm rms}$ the equipartion magnetic field strength 
$B_{\rm eq}=12\,\mu$G.
  
Overall the fits are quite good. The worst cases are WNB1328.3+6520 and WNB1458.0+4959, which happen to be the two most compact 
sources of our sample.
Apart from these two objects, we find that  $\nu_{b}$ is typically in the range 100 MHz - 1 GHz, while $t_{\rm OFF}/t_{s} > 0.1$
confirming that  these really are radio galaxies in which the fuelling of relativistic electrons has been in an off-state by a significant 
amount of time.
The spectra of WNB1328.3+6520 and WNB1458.0+4959 are steep power laws. According to the adopted model
we see the aged spectrum of the continuous injection phase. Thus, these sources may not yet be in the dying phase.

In general the radio cores are very weak, most having  $S_{\rm core}^{\rm 5~GHz}/S_{\rm tot}^{\rm 0.3~GHz}
 \le 3 \times 10^{-3}$.  The sources WNB1408.8+4630, WNB1438.0+3720 and WNB2317.4+4234 are different cases. 
Their low frequency spectrum is rather steep, but strongly flattens above 1.4 GHz, due to the presence of a 
bright core that is clearly detected in the 5 and 8.4\,GHz B-array images. These cores have a 
 flat spectrum characterized by $\alpha \le 0.5$ between these two frequencies and are likely still active.
In order to account for these components in the spectral fit, we added a power law to the CI$_{\rm OFF}$ model, using the observed 
spectral index and flux density as the normalization. In these three cases we may be observing fading lobes (produced by a previous
duty cycle), in conjunction with restarting activity in the core.

\subsection{Modelling the spectral index profiles}
\label{spixprof}
In three cases (WNB0951.9+6037, WNB1127.5+4927 and WNB1150.0+3749) the profile of the 0.3 to 1.4 GHz spectral index
can be traced all along the lobes. The traces are shown in the right-hand panels of Figs. 12 to 14.
Assuming a constant source expansion velocity the spectral index at a given distance from the core can be related to the synchrotron age of the 
electrons at that location. In particular, if the electrons were injected inside the core during the active phase, we expect that the break frequency
scales as:

\begin{equation}
\nu_{b}(d)=\frac{\nu_{b}(d_{\rm max})}{[(d/d_{\rm max})\cdot(1-t_{\rm OFF}/t_{s})+t_{\rm OFF}/t_{s}]^2}
\label{vbdist}
\end{equation}

where the distance $d$ ranges from $d=0$ at the core, up to  $d=d_{\rm max}$ at the edge of the source.
Thus, the break frequency along the lobes varies from a minimum of $\nu_{b}$, at $d=d_{\rm max}$,
 up to a maximum of $\nu_{b2}$, at $d=0$. The two limiting break frequencies $\nu_{b}$ and  $\nu_{b2}$
 are exactly the same as given in Eqs.\, (\ref{vb}) and  (\ref{vb1vb2}), respectively. 

WNB0951.9+6037 has a Z-shape and this complicates the interpretation of the spectral index
 profile: it is not possible to identify the regions where the electrons were injected during the active phase. 
On the other hand, in WNB1127.5+4927 and WNB1150.0+3749 the spectral index steepens going from the
 regions near the core outward. Thus, it is plausible that the electrons were injected near the core and 
 subsequently flowed outward. We fitted the spectral index trend in  WNB1127.5+4927
 with the CI$_{\rm OFF}$ model and obtained a minimum break frequency of $\nu_{b}\simeq (1800 \pm 250)$ MHz
 with $t_{\rm OFF}/t_{s}>0.79$. Such values are in very good agreement with those found from the
 fit of the global spectrum. In WNB1150.0+3749 the fit of the MJP model to the observed
spectral index trend gives $\nu_{b}\simeq (260 \pm 25)$ MHz and $t_{\rm OFF}/t_{s}=0.55\pm 0.05$.
For this source the CI$_{\rm OFF}$ model produces a significantly worse fit of the spectral trend than the MJP model,
in good agreement with the fit obtained for the integrated spectrum. 
We note that a source in which the activity had been stopped already for
a long time is expected to have $\nu_{b2}\simeq \nu_{b}$ and thus should have a rather constant spectral index
distribution. Strong spectral index variations along the lobes can be observed only 
if  $\nu_{b2}\gg \nu_{b}$, i.e. just after the injection has been switched off.

\subsection{Radiative ages}
\label{radages}
Assuming a constant magnetic field and no expansion after the switch-off, the total source age can be calculated from the break
frequency, $\nu_{b}$: 

\begin{equation}
t_{s}= 1590 \frac{B^{0.5}}{(B^2+B_{\rm IC}^2) [(1+z)\nu_{b}]^{0.5}}
\label{synage}
\end{equation} 

where the synchrotron age $t_{s}$ is in Myr, the magnetic field in $\mu$G, the break frequency $\nu_{b}$ in GHz,
while the inverse Compton equivalent field is $B_{\rm IC}=3.25(1+z)^{2}\,\mu$G.

By adopting the equipartition value $B_{eq}$ for the magnetic field strength we can thus
 derive the synchrotron age $t_{s}$. Finally, from the ratio $t_{\rm OFF}/t_{s}$, which 
is also given by the fit, we can determine the absolute durations of the active and relic phases, $t_{ci}$
 and $t_{\rm OFF}$.  The typical ages of the active phase are in the range $10^7~ -~ 10^8$ yrs.
The relic phase is typically shorter by an order of magnitude, but there are cases 
 in which the duration of the relic phase is comparable, if not longer, than that of the active phase
 (e.g. WNB1127.5+4927 and WNB1150.0+3749). It is quite hard to get a firm estimate on the relative ages of the
active and dying phases based purely on source statistics, as we would need unbiased and complete samples, constructed in the 
same way, of both dying and active sources.
However, we do have some general data on the identification of WENSS sources with nearby galaxies (de Ruiter et al., in preparation): preliminary
identification work gives us a sample of about 350 radio sources, extracted from the WENSS survey, if we follow the same criteria (except the
steepness of the radio spectrum, of course) as used for the
dying sources: identification with an elliptical galaxy with red magnitude $<17$, flux density at 327 MHz $>100$ mJy, and
unresolved at 327 MHz. The ratio of dying to active sources would then amount to roughly 2-3 \% (8/350), a bit less than the $\sim 10$ \%
implied by the synchrotron calculations given above. However, the condition that the source be unresolved 
(used by de Breuck et al. \cite{debreuck00} in their construction of a sample of steep spectrum sources) provides un unknown but probably 
significant bias against finding dying sources, since these are expected to be on average older and more extended than younger,
still active, sources. Therefore the discrepancy by a factor of 4 between the ratio of dying to active sources estimated from source statistics
as compared to the value obtained from synchrotron ages, is by no means worrying.

All the ages are given in Table \ref{tab:spectralpar}, together with the best fit parameters.

It is interesting to compare the radiative ages of the dying sources with
those of radio galaxies with similar luminosities in the low frequency range, that are still in the continuous injection phase.
Data on active radio galaxies were taken from the sample of B2 radio galaxies (Parma et al.  \cite{Parma99}). 
For these sources we have recomputed the equipartition and the radiative ages following the procedure of Sect. \ref{equip}.
The dying sources indeed tend to be older than the B2 radio galaxies. In fact, the median age ($t_{s}$)  of the dying sources is
63 Myr, against 16 Myr of the B2 radio galaxies; all of the dying sources are older than
16 Myr, the median age of the B2 radio galaxies. Even if we consider instead the median duration of the active phase of the
dying sources in our sample, we find that this is 48 Myr, three times the age of the B2 sources. These results are consistent with
the scenario in which the sources dubbed "dying" indeed represent the last stage in the lives of radio sources.

\subsection{The environment of the dying sources}
\label{ambient}
We investigated the gaseous environment of the dying radio galaxies by checking for diffuse X-ray emission in the 
Rosat All Sky Survey (RASS). We extracted the 0.1 - 2.4 keV image of a $10\arcmin \times 10\arcmin$ field around
each radio source. The X-ray images were corrected for the background and smoothed with a 
$\sigma= 45\arcsec$ Gaussian kernel.

The RASS count-rate contours are shown in Figs. \ref{fig:fig16} and \ref{fig:fig17}, overlayed on the  1.4\,GHz VLA B-array images.

Cross-correlation of the NED database with the RASS X-ray images gives the following results:
\label{xraynotes}
\begin{itemize}

\item WNB0722.4+4129. The radio source is located near the centre of the Abell cluster A\,580, which has also been detected in the RASS.

\item WNB0951.9+6037. The RASS image suggests the presence of a faint X-ray source coinciding 
with the dying radio source, but no clusters are known within 10\arcmin\, from the radio galaxy.
 
\item WNB1023.3+2959. No X-ray emission is seen down to the sensitivity level of the RASS; no cluster is 
known within a 10\arcmin\, radius from the radio galaxy.

\item WNB1127.5+4927. No X-ray emission is seen down to the sensitivity level of the RASS; no cluster is 
known within a 10\arcmin\, radius from the radio galaxy.  

\item WNB1150.0+3749. No X-ray emission is seen down to the sensitivity level of the RASS; NED reports a cluster 
(NSCJ115245+373321) at 1\farcm 9 from the radio galaxy, but the photometric redshift of the cluster is
z=0.1456, while the radio galaxy has a spectroscopic redshift of z=0.228.

\item WNB1257.4+3137. A faint X-ray source is present in the RASS image at the position of the dying galaxy.
NED lists a cluster (NSCJ125947+312215) at  1\farcm 7 from the radio source, with a photometric redshift of z= 0.0585; this is in agreement
with the spectroscopic redshift of the radio source, z=0.05172.

\item WNB1328.3+6520. An extended X-ray source is present in the RASS image, a few arcminutes north of the radio source position. 
The nearest cluster, Zw\,1328.9+6518, lies 4\arcmin\, south-east of the radio source. It is therefore not
clear whether the X-ray emission is associated with this cluster.

\item WNB1408.8+4630. Faint  X-ray emission is present in the RASS image, at the position of the radio source. The X-ray emission
is probably associated with the cluster NSCJ141045+461634 (photometric redshift of z=0.178). However the redshift of the radio source
is 0.13255.
 
\item WNB1438.0+3720.  The radio source belongs to the cluster MACSJ1440+3707, which has been detected in the RASS.

\item WNB1458.0+4959.  The radio source is located in the cluster Abell A\,2011. The RASS image shows a peak of X-ray emission
in correspondence with the radio source and diffusion emission about 3\arcmin\, to the east. It is unclear whether the source is
located at the cluster centre or in the periphery;

\item WNB2317.4+4234. The RASS image shows strong diffuse X-ray emission, confirmed by a Chandra observation (see bottom-right
panel of Fig.  \ref{fig:fig13}), at the position of the radio source
and coincident with the galaxy NGC\,7618.
\end{itemize} 

No firm conclusions can be drawn about the environment of dying sources, due to the small number of objects involved.
Although about half of the dying sources are located in clusters, there are some (WNB1023.3+2959, WNB1150.0+3749 and  WNB1127.5+4927) 
which appear to be isolated.

\section{Conclusions and Summary}
\label{sec:summary}

The definition of dying source is based on spectral shape and the absence (or at least weakness) of a radio core powering large-scale jets. 
Using this definition
our search for dying sources started with eleven candidates, which in the end revealed six truly dying sources, and three more in
which a radio core detected at higher frequencies suggests activity that has recently restarted. 

We emphasize that this is the total number of dying sources present in the entire WENSS catalogue, at least of those with angular sizes
less than about one arcmin, flux density at 1.4 GHz $>10$ mJy in the NVSS catalogue, and identified with galaxies with $m_r<17$.
Because of these restrictions we may very well have missed sources with {\it very} steep spectra (i.e. those with a 20 cm flux below the adopted 10 mJy
limit) and also the more extended sources. We are constructing a new sample of steep spectrum sources, always using WENSS, NVSS and an
optical magnitude limit of 17 in the red, but this time without any restrictions on angular size restriction and 20 cm flux density 
(de Ruiter et al. in prep.). Until then it is very hard to give an estimate of the frequency of dying sources in general radio catalogues, and in 
particular a statistical estimate of the relative time scales of the dying and active phases, to be compared with the $t_{\rm OFF}/t_{s}$ directly 
obtained by the model fits of the radio spectra (see Table \ref{tab:spectralpar}). 

The typical ages of the active phase derived by the spectral fits are in the range $10^7~ -~ 10^8$ yrs.
The relic phase is typically shorter by an order of magnitude, but there are cases 
in which its duration is comparable, if not longer, than that of the active phase
(e.g. WNB1127.5+4927 and WNB1150.0+3749).
A dying source tends to be older than a general B2 radio galaxy. In fact, the median age $t_{s}$ of the dying sources is
63 Myr, against 16 Myr of the B2 radio galaxies; all of the dying sources discussed here are older than
these 16 Myr. 
As we argued in the previous section, it is difficult to get an estimate of the relative ages from source
statistics, due to the present lack of well defined samples of both dying and still active sources. Preliminary data suggest that
synchrotron ages predict about four times as many dying sources than actually found, but if we take into account the possible
biases in our search procedure, plus the intrinsic uncertainties in the synchrotron calculations, we do not consider this a serious
discrepancy.


Two of the sources are only marginally resolved at our best resolution. Their linear sizes are $\le$ 10 kpc, 
so they formally belong to the class of {\it Compact Steep-Spectrum Sources}, although they are less 
powerful than the classical members found in the 3CR and PW catalogues.
  
Although about half of the dying sources are located in clusters, there are some which appear to be isolated. 
Note that the three dying sources found in the WENSS minisurvey sample (de Ruiter et al. \cite{deruiter98})
are all in a cluster (Murgia et al. \cite{Murgia05}). Considering the small numbers involved, no firm conclusions can be drawn about 
the environment of dying sources.

The radio luminosities at 325 MHz are close to the break in the luminosity function. 
According to the spectral evolution scenario we have assumed that the radio power before the switch-off of the central engine was 
somewhat higher and has decreased due to the radiation losses, but by no more than a factor 2--3. 


\begin{acknowledgements}
The National Radio Astronomy Observatory is operated by Associated Universities, Inc., under contract with
the National Science Foundation.
This research made use of the NASA/IPAC Extragalactic Database (NED) which is operated 
by the Jet Propulsion Laboratory, California Institute of Technology, under contract with the National
Aeronautics and Space Administration.
This research made use also of the CATS Database (Astrophysical CATalogs support System).
The optical DSS2 red images were taken from: http://archive.eso.org/dss/dss.
We thank the staff of the National Galileo Telescope (TNG) for carrying out the optical observations in service mode.
Finally we thank Emanuela Orr{\`u}, who gave us valuable advice for the reduction of the low-frequency data.
\end{acknowledgements}

\clearpage
\begin{table*}
\scriptsize
\caption[]{Observing log}
\label{tab:observing log}
\begin{flushleft}
\begin{tabular}{lccrcrc}
\hline\noalign{\smallskip}
Name                     & Array & frequency & Obs. Time & beam          & P.A.        & noise  \\
                              &           &  GHz         &   min.        & \arcsec        &  \degr & mJy/beam \\
\noalign{\smallskip}
\hline\noalign{\smallskip}

WNB0722.4+4129 & B     & 1.425    &  69        &  4.32 $\times$ 3.69   & 8         & 0.042 \\
                               & B     & 4.860    &   8        &  1.21 $\times$ 1.06    & $-11$  & 0.046 \\
                               & B     & 8.460    &   8        &  0.68 $\times$ 0.61    & $-17$  & 0.022 \\
                               & D     & 4.860    &   8        & 13.52 $\times$ 11.83 & $-21$  & 0.021 \\
                               & D     & 8.460    &   8        &  7.86 $\times$ 6.91    & $-22$  & 0.014 \\
WNB0951.9+6037 & A     & 0.325    &  53        &  6.11 $\times$ 4.62    & 4         & 0.630  \\
                               & B     & 1.425    &  80        &  4.84 $\times$ 3.64   & $-5$    & 0.026 \\                
                               & B     & 4.860    &   8         &  1.46 $\times$ 1.09   & 38       & 0.047 \\
                               & B     & 8.460    &   8        &  0.80 $\times$ 0.61    & 37       & 0.029 \\
                               & D     & 4.860    &   8        & 15.08 $\times$ 11.46 & 25       & 0.020 \\
                               & D     & 8.460    &   8        &  8.66 $\times$ 6.47    & 25       & 0.017 \\  
WNB1023.3+2959 & B     & 1.425    &  80        &  4.42 $\times$ 3.84   & 38       & 0.024 \\
                               & B     & 4.860    &   8        &  1.25 $\times$ 1.17    & 41       & 0.045 \\
                               & B     & 8.460    &   9        &  0.69 $\times$ 0.66    & 58       & 0.025 \\
                               & D     & 4.860    &   8        & 14.03 $\times$ 12.29 & 51       & 0.013 \\
                               & D     & 8.460    &   8        &  8.07 $\times$ 7.17    & 49       & 0.011 \\
WNB1127.5+4927 & A     & 0.325    &  73        &  6.65 $\times$ 4.87    & 6        & 0.700  \\
                               & B     & 1.425    &  75        &  4.47 $\times$ 3.66   & 0        & 0.029 \\
                               & B     & 4.860    &   8        &  1.41 $\times$ 1.11    & 54       & 0.046 \\
                               & B     & 8.460    &   9        &  0.78 $\times$ 0.63    & 55       & 0.029 \\
                               & D     & 4.860    &   8        & 14.36 $\times$ 11.74 & 42       & 0.018 \\
                               & D     & 8.460    &   8        &  7.98 $\times$ 6.66    & 27       & 0.016 \\
WNB1150.0+3749 & A     & 0.325    &  60        &  5.46 $\times$ 4.91    & $-1$    & 0.700 \\
                               & B     & 1.425    &  50        &  4.56 $\times$ 3.89   & 53       & 0.034 \\
                               & B     & 4.860    &   9        &  1.28 $\times$ 1.14    & 46       & 0.045 \\
                               & B     & 8.460    &   8        &  0.70 $\times$ 0.64    & 48       & 0.027 \\
                               & D     & 4.860    &   8        & 13.99 $\times$ 12.39 & 50       & 0.016 \\
                               & D     & 8.460    &   8        &  7.79 $\times$ 7.09    & 30       & 0.015 \\
WNB1257.4+3137 & B     & 1.425    &  69        &  4.59 $\times$ 4.11    & $-64$ & 0.028 \\
                               & B     & 4.860    &   7        &  1.28 $\times$ 1.19    & 60       & 0.045 \\
                               & B     & 8.460    &   7        &  0.71 $\times$ 0.32    & 73       & 0.025 \\
                               & D     & 4.860    &   8        & 14.40 $\times$ 13.07 & 79       & 0.020 \\
                               & D     & 8.460    &   8        &  9.02 $\times$ 7.34    & 71       & 0.012 \\
WNB1328.3+6520 & B     & 1.425    &  49        &  5.35 $\times$ 3.65   & 43       & 0.060 \\
                               & B     & 4.860    &   8        &  1.47 $\times$ 1.05    & 26       & 0.049 \\
                               & B     & 8.460    &   8        &  0.83 $\times$ 0.61    & 30       & 0.030 \\
                               & D     & 4.860    &   8        & 17.94 $\times$ 11.50 & 56       & 0.020 \\
                               & D     & 8.460    &   8        &  9.73 $\times$ 6.89    & 43       & 0.022 \\
WNB1408.8+4630 & B     & 1.425    &  58        &  4.36 $\times$ 3.72    & 29       & 0.050 \\
                               & B     & 4.860    &   8        &  1.24 $\times$ 1.05    & 1         & 0.045 \\
                               & B     & 8.460    &   8        &  0.69 $\times$ 0.61    & $-11$  & 0.028 \\
                               & D     & 4.860    &   8        & 15.34 $\times$ 12.00 & 64       & 0.018 \\
                               & D     & 8.460    &   8        &  8.45 $\times$ 6.99    & 56       & 0.017 \\
WNB1438.0+3720 & B     & 1.425    &  80        &  4.36 $\times$ 3.72   & 31       & 0.032 \\
                               & B     & 4.860    &   8        &  1.20 $\times$ 1.07    & 6         & 0.044 \\
                               & B     & 8.460    &   8        &  0.68 $\times$ 0.61    & 14       & 0.026 \\
                               & D     & 4.860    &   8        & 14.48 $\times$ 12.20 & 65       & 0.018 \\
                               & D     & 8.460    &   8        &  8.06 $\times$ 7.18    & 61       & 0.016 \\ 
WNB1458.0+4959 & B     & 1.425    &  58        &  4.44 $\times$ 3.67    & 3        & 0.050 \\
                               & B     & 4.860    &   8        &  1.31 $\times$ 1.03    & $-20$  & 0.046 \\
                               & B     & 8.460    &   7        &  0.73 $\times$ 0.60    & $-24$  & 0.030 \\
                               & D     & 4.860    &   8        & 16.77 $\times$ 12.03 & 73       & 0.022 \\
                               & D     & 8.460    &   9        &  9.26 $\times$ 6.77    & 65       & 0.017 \\
WNB2317.4+4234 & A     & 0.325    &  35        &  5.24 $\times$ 4.62    & 35       & 1.000 \\
                              & B     & 1.425    &  55        &  4.29 $\times$ 3.66    & 22       & 0.032 \\
                              & B     & 4.860    &   9        &  1.23 $\times$ 1.02     & 20       & 0.035 \\
                              & B     & 8.460    &   9        &  0.71 $\times$ 0.71     & 0         & 0.037 \\
                              & D     & 4.860    &  14        & 13.62 $\times$ 11.79 & 29       & 0.016 \\
                              & D     & 8.460    &  14        &  7.61 $\times$   6.77  & 10       & 0.015 \\
\noalign{\smallskip}
\hline
\end{tabular}
\end{flushleft}
\end{table*}

\clearpage
\begin{table*}
\caption[]{Observational data}
\label{tab:observational  data}
\begin{flushleft}
\begin{tabular}{lccrclrccc}
\noalign{\smallskip}
\hline\noalign{\smallskip}

\multicolumn{1}{c}{Name} & RA & DEC               & redshift & $ m_{r}$ & frequency & S & \multicolumn{3}{c}{Size} \\
                                           & \multicolumn{2}{c}{(J2000)} &              &                &                  & &  FWHM  & P.A. & LAS \\
                                           & $^h\,  ^m\, ^s $                & $\degr$\, \arcmin \, \arcsec    &                &          & GHz &  mJy & \arcsec  & $\degr$  &  \arcsec \\

\noalign{\smallskip}
\hline\noalign{\smallskip}
WNB0722.4+4129 & 07 25 57.22 & 41 23 05.6 & 0.11183    & 14.9 &  1.4      & 32.3  &               &          & 19 $\times$ 13 \\
\noalign{\smallskip}
\hline\noalign{\smallskip}
WNB0951.9+6037 & 09 55 29.93 & 60 23 17.2 & 0.19908    & 15.9  & 0.33     &  99   &               &          &  72     \\
central comp.  & 09 55 29.87 & 60 23 17.2 &            &       & 0.33     &  6.1  & 2.3 $\times$ 0.0     & 132 \\                
               &             &            &            &       & 1.4      &  1.8  & 2.9 $\times$ 1.1     &  80  \\
               &             &            &            &       &  8.4     & 0.18  & 0.45 $\times$ 0.00   &  35  \\
N lobe         &             &            &            &       &  0.33    &  46   &               &           & 45 $\times$ 10 \\
               &             &            &            &       &  1.4     &  3.2  &               &           & 31 $\times$ 10 \\
S lobe         &             &            &            &       &  0.33    &  47   &               &           & 34 $\times$ 10 \\
               &             &            &            &       &  1.4     &  3.8  &               &           & 34 $\times$ 10 \\
\noalign{\smallskip}
\hline\noalign{\smallskip}
WNB1023.3+2959 & 10 26 11.87 & 29 43 45.5 & 0.24228    & 17.0  &  1.4     &  11   &               &           & 23 $\times$ 14 \\
\noalign{\smallskip}
\hline\noalign{\smallskip}

WNB1127.5+4927 & 11 30 18.36 & 49 11 15.4 & 0.25965    & 16.3  &  0.33    & 132   &               &           & 53 $\times$ 18 \\
E lobe         &             &            &            &       &  0.33    &  50   &               &           & 25 $\times$ 13 \\
               &             &            &            &       &  1.4     & 6.7   &               &           & 25 $\times$ 13 \\
W lobe         &             &            &            &       &  0.33    &  72   &               &           & 20 $\times$ 18 \\
               &             &            &            &       &  1.4     & 11    &               &           & 16 $\times$ 16 \\
\noalign{\smallskip}
\hline\noalign{\smallskip}

WNB1150.0+37409& 11 52 36.50 & 37 32 43.7 & 0.22848    & 16.0  &  0.33    & 287   &               &           & 31 $\times$ 25 \\    
               &             &            &            &       &  1.4     & 15    &               &           & 30 $\times$ 22 \\
\noalign{\smallskip}
\hline\noalign{\smallskip}

WNB1257.4+3137 & 12 59 51.99 & 31 21 06.5 & 0.05172    & 13.6  &  1.4      &  32  &               &           & 52 $\times$ 3 0 \\
central comp.  & 12 59 52.00 & 31 21 06.2 &            &       &  5.0      & 0.59 & 2.9 $\times$ 0.0     & 109       &         \\
\noalign{\smallskip}
\hline\noalign{\smallskip}

WNB1328.3+6520 & 13 30 02.77 & 65 04 59.3 & 0.21899    & 15.9  &  1.4      & 10   & 2.7 $\times$ 1.6     & 167       &        \\
               &             &            &            &       &  5.0      &  1.7 & 0.78 $\times$ 0.00   & 87        &         \\
               &             &            &            &       &  8.4      &  1.0 & 0.17 $\times$ 0.00   & 169       &         \\
\noalign{\smallskip}
\hline\noalign{\smallskip}

WNB1408.8+4630 & 14 10 48.18 & 46 15 57.5 & 0.13255    & 15.4  &  1.4      & 11   &               &           & 18      \\
central comp.  & 14 10 48.19 & 46 15 57.7 &            &       &  5.0      &  5.3 & 0.28 $\times$ 0.14   & 102       &         \\
               &             &            &            &       &  8.4      &  4.7 & 0.1 $\times$ 0.0     & 144       &         \\
\noalign{\smallskip}
\hline\noalign{\smallskip}

WNB1438.0+3720 & 14 40 03.45 & 37 07 27.5 & 0.09661    & 15.3  &  1.4      & 15.0 &               &           & 30 $\times$ 12  \\
central comp.  & 14 40 03.45 & 37 07 27.4 &            &       &  5.0      & 0.86 & 0.68 $\times$ 0.00   &  78       &         \\
               &             &            &            &       &  8.4      & 0.71 & 0.68 $\times$ 0.16   &  74       &         \\
\noalign{\smallskip}
\hline\noalign{\smallskip}

WNB1458.0+4959 & 14 59 43.24 & 49 47 15.9 & 0.16782    & 15.8  &  1.4      & 15   &  2.9 $\times$ 1.6    &  108      &         \\  
               &             &            &            &       &  5.0      &  1.9 &  2.1 $\times$ 1.1    & 113       &         \\
\noalign{\smallskip}
\hline\noalign{\smallskip}

WNB2317.4+4234 & 23 19 47.23 & 42 51 09.5 & 0.01730    & 12.5  &  1.4      & 27   &               &           & 48 $\times$ 30 \\
head           & 23 19 47.23 & 42 51 09.2 &            &       & 0.33      & 23   &  5.9 $\times$ 2.6    & 122       &         \\
               &             &            &            &       &  1.4      & 8.1  & 2.3 $\times$ 1.9     &  163      &         \\
               &             &            &            &       &  5.0      & 3.3  & 0.35 $\times$ 0.27   & 149       &         \\
               &             &            &            &       &  8.4      & 2.5  & 0.31 $\times$ 0.09   &  7        &         \\
\noalign{\smallskip}
\hline
\end{tabular}
\end{flushleft}
\end{table*}

\clearpage
\begin{table*}
\caption[]{Additional redshifts}
\label{tab:redshifts}
\begin{flushleft}
\begin{tabular}{cccc}
\hline\noalign{\smallskip}
Name & RA(J2000) & DEC(J2000) & Redshift \\
          & $^h\,  ^m\, ^s$       & $\degr$\, \arcmin \, \arcsec \\
\noalign{\smallskip}
\hline\noalign{\smallskip}

WNB0951.9+6037 & 09 55 31.32 & 60 23 15.3 & 0.20603 \\
WNB1150.5+3749 & 11 52 35.97 & 37 32 51.7 & 0.22867 \\
WNB1438.0+3720 & 14 40 03.83 & 37 06 55.0 & 0.09483 \\
WNB1458.0+4959 & 14 59 43.44 & 49 47 35.2 & 0.16545 \\

\noalign{\smallskip}
\hline
\end{tabular}
\smallskip

\end{flushleft}
\end{table*}

\clearpage
\begin{table*}
\caption[]{Integrated flux densities}
\label{tab:integrated flux}
\begin{flushleft}
\begin{tabular}{lrrrrrrrrrl}
\hline\noalign{\smallskip}
Name & $ S_{74}$ & $ S_{151}$ & $ S_{232}$ & $ S_{325}$ &
 $S_{365}$ &
$ S_{408}$ & $ S_{1.4}$ & $ S_{4.8}$ &
$ S_{8.4}$ & References\\
    & mJy & mJy &  mJy & mJy &mJy &mJy & mJy &mJy & mJy & \\
\noalign{\smallskip}
\hline\noalign{\smallskip}
WNB0722.4+4129 & 1180 &  655 &     &  272 &     &  170 & 35.5 & 4.18 & 1.35 & 5,8 \\

WNB0951.9+6037 & 419 &  290 & 170 &  103 &     &      & 10.9 & 0.40 & 0.12 & 2,6 \\

WNB1023.3+2959 & - &  240 &     &   81 &     &      & 12.0 & 0.80 & 0.26 & 4 \\

WNB1127.5+4927 & 495 &  364 & 228 &  132 &     &  113 & 17.6 & 0.28 &$<0.09$ & 5,6,8 \\

WNB1150.0+3749 & 4600 & 1582 & 620 &  414 & 341 &  250 & 17.1 & 0.43 & $<0.08$ & 5,6,7,8 \\

WNB1257.4+3137 & 792 &  540 &     &  268 &     &      & 34.0 & 3.29&  1.09  & 4 \\

WNB1328.3+6520 & 369 &      &     &   74 &     &      & 10.5 & 1.72&  1.09  & \\

WNB1408.8+4630 & 527 &  467 &     &  103 &     &      & 14.2 & 6.51 & 5.30  & 5 \\

WNB1438.0+3720 & 973 &  550 &     &  219 &     &      & 16.7 & 1.56 & 0.86 & 1 \\

WNB1458.0+4959 & 1120 &  670 &     &  175 &     &      & 18.7 & 2.82 & 1.14 & 5 \\

WNB2317.4+4234 & 1218 & 1190 & 330 &  321 &     &  260 & 40.2 & 4.25 & 3.01 & 3,6,8 \\

\noalign{\smallskip}
\hline
\end{tabular}
\smallskip

Data at 4.8 GHz and 8.4 GHz are from the present paper; data at 325 MHz and 1.4 GHz are from  De Breuck et al. 
(\cite{debreuck00}). References for other frequencies:

 1-Hales et al. (\cite{Hales88}): 6CII, 151 MHz

 2-Hales et al. (\cite{Hales90}): 6CIII, 151 MHz

 3-Hales et al. (\cite{Hales93}): 6CVI, 151 MHz 

 4-Waldram et al. (\cite{Wal96}): 7C4, 151 MHz

 5-Riley et al. (\cite{Riley99}): 7C, 151 MHz

 6-Zhang et al (\cite{Zhang97}): Miyun, 232 MHz

 7-Douglas et al. (\cite{Douglas96}): TXS, 365 MHz

 8-Ficarra et al. (\cite{Fic85}): B3.1, 408 MHz 
\end{flushleft}
\end{table*}

\clearpage
\begin{table*}
\caption[]{Intrinsic parameters}
\label{tab:intrinsic parameters}
\begin{flushleft}
\begin{tabular}{lccccll}
\hline\noalign{\smallskip}
Name & redshift & $M_r$ & $\log(P_t/0.33 MHz)$ & Linear Size & $u_{min}$                      & $ B_{eq}$     \\
           &              &            & WHz$^{-1}$                &   kpc            & $10^{-11}$ erg~ cm$^{-3}$ & $\mu$G \\
\noalign{\smallskip}
\hline\noalign{\smallskip}
WNB0722.4+4129 & 0.11183  & $-23.6$ &  24.93              & 38 $\times$ 26  & 1.9 & 15.2 \\
\noalign{\smallskip}
\hline\noalign{\smallskip}
WNB0951.9+6037 & 0.19908 & $-24.0$ &   25.05              & 234 \\
N lobe                    &               &              &   24.71              & 147 $\times$ 33 & 0.6 & 8.5 \\
S lobe                    &               &              &   24.72              & 111 $\times$ 33 & 0.6 & 9.1 \\
central comp.         &               &              &   23.84              & 7.5  \\
\noalign{\smallskip}
\hline\noalign{\smallskip}
WNB1023.3+2959 & 0.24228 & $ -23.4$ &   25.15             & 87 $\times$ 53 & 0.6 & 8.7 \\
\noalign{\smallskip}
\hline\noalign{\smallskip}
WNB1127.5+4927 & 0.25965 & $-24.3$ &   25.43               & 211 $\times$ 72 \\
E lobe         &         &       &   25.01               & 100 $\times$ 52& 0.6 & 8.5 \\
W lobe         &         &       &   25.17               &  80 $\times$ 72& 0.6 & 8.3 \\
\noalign{\smallskip}
\hline\noalign{\smallskip}
WNB1150.0+3749 & 0.22848 & $-24.3$ &   25.8                & 112 $\times$ 91& 1.1 & 12 \\
\noalign{\smallskip}
\hline\noalign{\smallskip}
WNB1257.4+3137 & 0.05172 & $-23.2$ &   24.22               &  52 $\times$ 30& 0.4 & 7.0  \\
\noalign{\smallskip}
\hline\noalign{\smallskip}
WNB1328.3+6520 & 0.21899 & $-24.3$ &   25.01               & 0.60    & 0.9 & 33.6 \\
  \noalign{\smallskip}
\hline\noalign{\smallskip}
WNB1408.8+4630 & 0.13255 & $-23.5$ &   24.67               & 42 \\
\noalign{\smallskip}
\hline\noalign{\smallskip}
WNB1438.0+4959 & 0.09661 & $-22.9$ &   24.70               & 53 $\times$ 23 & 1.5 & 13.6 \\
 \noalign{\smallskip}
\hline\noalign{\smallskip}
\noalign{\smallskip}
WNB1458.0+4959 & 0.16782 & $-23.7$ &   25.12               & 6.0 $\times$ 3.1 & 1.6 & 44.8 \\
\noalign{\smallskip}
\hline\noalign{\smallskip}
WNB2317.4+4234 & 0.01730 & $-21.0$ &   22.99               & 11 $\times$ 7 \\

\noalign{\smallskip}
\hline
\end{tabular}
\smallskip
\end{flushleft}
\end{table*}

\clearpage
\begin{table*}
\caption[]{Spectral parameters}
\label{tab:spectralpar}
\begin{flushleft}
\begin{tabular}{rrrrrrrrc}
\hline\noalign{\smallskip}
Name            & $\chi^2_{red.}$&$\nu_{b}$&$t_{\rm OFF}/t_s$&$\nu_{b2}$& $t_s$ & $t_{ci}$&$t_{\rm OFF}$ & Notes  \\
               &          &  MHz   &       &  MHz     &  Myr  &  Myr &  Myr & \\
\noalign{\smallskip}
\hline\noalign{\smallskip}
WNB0722.4+4129 &  1.2     & $228^{+13}_{-10}$    &  $0.12^{+0.04}_{-0.03}$  & 15833  &  50 &  44 & 6.0 &   \\
\noalign{\smallskip}
\hline\noalign{\smallskip}
WNB0951.9+6037 &  1.7    &  $415^{+53}_{-29}$    &  $0.35^{+0.05}_{-0.13}$  &  3387 & 68  & 44 & 24 &          \\
\noalign{\smallskip}
\hline\noalign{\smallskip}
WNB1023.3+2959 &  5.9    &  $<105$               &  $0.10^{+0.01}_{-0.01}$&  $<10500$ &  $>130$   &  $>120$ & $>13$&     \\
\noalign{\smallskip}
\hline\noalign{\smallskip}
WNB1127.5+4927 &  2.5    & $1352^{+50}_{-721}$    &  $>0.53$   &  $<4813$ & 36 &  $<17$ & $>19$  &  (1)      \\
\noalign{\smallskip}
               &  0.81   & $1832^{+267}_{-241}$     &  $>0.79$ &  $<2935$ & 32     &  $<7.0$ & $>25$&  (2)        \\
\hline\noalign{\smallskip}
WNB1150.0+3749 &  1.3    & $30^{+2}_{-2}$ & $0.20^{+0.08}_{-0.02}$  & 383 & 170 & 120 & 48   &    (3)      \\
\noalign{\smallskip}
               &  2.4    & $260^{+28}_{-26}$ & $0.55^{+0.06}_{-0.05}$   & 1040 & 58 & 29 & 29  &  (4)          \\
\noalign{\smallskip}
\hline\noalign{\smallskip}
WNB1257.4+3137 &  4.3    & $390^{+27}_{-26}$     &  $0.19^{+0.02}_{-0.02}$   & 10803    & 110 & 87 & 20 &   \\
\noalign{\smallskip}
\hline\noalign{\smallskip}
WNB1328.3+6520 & 10.6    & $61^{+6}_{-6}$    & $0.02^{+0.01}_{-0.01}$ & $>20000$ & 29 & 28 &0.6 &  \\
\noalign{\smallskip}
\hline\noalign{\smallskip}
WNB1408.8+4630 & 0.5     & $<144$    &   $-$ &    $-$    & $-$& $-$&  \\
\noalign{\smallskip}
\hline\noalign{\smallskip}
WNB1438.0+4959 & 2.9     & $195^{+129}_{-11}$    & $0.24^{+0.08}_{-0.07}$   & 3385      &  63  &  48 &15 &   \\
 \noalign{\smallskip}
\hline\noalign{\smallskip}
\noalign{\smallskip}
WNB1458.0+4959 & 20.0    & $34^{+2}_{-2}$ & $0.043^{+0.004}_{-0.004}$& $>20000$ & 26 & 25 & 1.1 &  \\
\hline\noalign{\smallskip}
\noalign{\smallskip}
WNB2317.4+4234 & 7.2     & $216^{+289}_{-115}$ & $0.25^{+0.18}_{-0.09}$ & 3456& $-$&$-$& $-$ &   \\
\noalign{\smallskip}
\hline
\end{tabular}
\smallskip

Notes:\\
(1) WNB1127.5+4927; Integrated $CI_{OFF}$ model.\\
(2) WNB1127.5+4927; Sp. index profiles, JP model.\\
(3) WNB1150.0+3749; Integrated MJP model with $B/B_{\rm IC}$=2.4 fixed (i.e. $B=12\mu G$).\\
(4) WNB1150.0+3749; Sp. index profiles, JP model with $B/B_{\rm IC}$=2.4 fixed (i.e. $B=12\mu G$).\\
\end{flushleft}
\end{table*}

\end{document}